\documentclass[]{aa}
\usepackage{graphicx}
\usepackage{natbib}
\bibpunct{(}{)}{;}{a}{}{,}
\usepackage{txfonts}
\begin{document}

\title{A Multiplicity Survey of the $\rho$ Ophiuchi Molecular Clouds
      \thanks{Based on observation with the New Technology
             Telescope (NTT, proposals 65.I-0067 and 67.C-0354) 
             and the 3.6~m telescope (proposal 65.I-0086) at the 
             European Southern Observatory (ESO), La Silla, 
             Chile and the 3.5~m telescope at Calar Alto, Spain.
             }
      }

\author{Th.~Ratzka\inst{1}
       \and
       R.~K\"ohler\inst{1,2}
       \and
       Ch.~Leinert\inst{1}
       }

\offprints{Thorsten Ratzka Max-Planck-Institute for Astronomy (MPIA), \email{ratzka@mpia.de}}

\institute{Max-Planck-Institute for Astronomy (MPIA), 
          K\"onigstuhl 17, 69117 Heidelberg, Germany
	  \and
	  Sterrewacht Leiden, Niels Bohrweg 2, 2300 RA Leiden, The Netherlands
          }

\date{Received 1 October 2004 / Accepted 22 March 2005}

\abstract{We present a volume-limited multiplicity survey with magnitude cutoff ($\mathrm{m}_\mathrm{K}\leq 10.5\mathrm{\ mag}$) of 158 young stellar objects located within or in the vicinity of the $\rho$ Ophiuchi Dark Cloud. With exception of eleven already well observed objects, all sources have been observed by us in the K-band with 3.5~m telescopes by using speckle techniques. The separation range covered by our survey is $0.13''\leq\theta\leq 6.4''$, where the lower limit is given by the diffraction limit of the telescopes and the upper limit by confusion with background stars. The multiplicity survey is complete for flux ratios $\geq 0.1$ ($\Delta\mathrm{m}_\mathrm{K}\leq 2.5$) at the diffraction limit. After taking the background density into account the degree of multiplicity is $29.1\%\pm 4.3\%$ and thus only marginally higher than the value $23.5\%\pm 4.8 \%$ derived for the given separation range for the main-sequence solar-like stars in the solar neighbourhood \citep{dm}. We discuss the implications of these findings.
 
         }

\maketitle


\section{Introduction}
The detection of an overabundance by a factor of two of binaries among the young stars in Taurus when compared to the results for the main sequence \citep{ghez, leinert93, rz} made it very clear that binarity indeed is the dominant mode of star formation. Consequently, in the years after these studies both theoretical and observational work on binaries among young stars and binary formation was intensified. Observationally, two main routes were followed: studying the fraction of binaries in associations and in young clusters, both with the aim  to learn about the conditions which influence the preference of binary over single star formation. The study of associations, all of which were about at the same distance of $\approx 150~\mathrm{pc}$ and are about equally young (several million years) has so far not given a clear picture, see e.g. the summary by \cite{duchene}: the duplicity is high in Taurus (see references above), CrA \citep{ghez2, rz} and Scorpius \citep{koehler2}, while it almost corresponds to main-sequence values or is even lower in Chamaeleon and Lupus \citep{brandner, rz, koehler3}. To identify the reason for this different behaviour will need further observational studies and continued discussions on the interpretation, although \cite{durisen} proposed a possible explanation, namely that fragmentation should lead to lower fractions of binaries for higher initial cloud temperature. 

The situation is somewhat more settled in the case of clusters. The advantage here is that clusters of different age can be studied in order to get information on the temporal evolution. The result is that even the youngest of them, the Trapezium, does not show an overabundance of binaries \citep{prosser, petr, padgett}. It is true that N-body simulations, e.g. \cite{kroupa} indicate that in dense clusters the fraction of binaries could be reduced by gravitational interactions within 1 million of years from `high' to `normal'. But the assumption that the lower fraction of binaries in dense clusters may be intrinsic and determined by the density as a parameter remains an attractive hypothesis \citep{duchene2}.

Although there is evidence for an overabundance of multiple systems in the Ophiuchus star forming region compared to the main-sequence \citep{ghez2, sgl, duchene}, the statistics are based only on a small number of systems observed with various techniques. To derive a survey for the $\rho$ Ophiuchi cloud complex (section~2) that is comparable to our surveys of the Taurus star forming region \citep{leinert93, koehler1} we created a magnitude-limited sample (section~3) based on previous work that determined the cloud membership of our targets. With exception of some well-studied objects we observed our complete survey by using speckle techniques at 3.5~m telescopes (section~4). To reduce the data we used a software package (section~5) developed in our group during the last years. After the correction of the raw data (section~6) we discuss the results in terms of age and density effects (section~7). A summary of the results is given in section~8. 

\begin{figure*}
\centering
\includegraphics[height=18.1cm,angle=270]{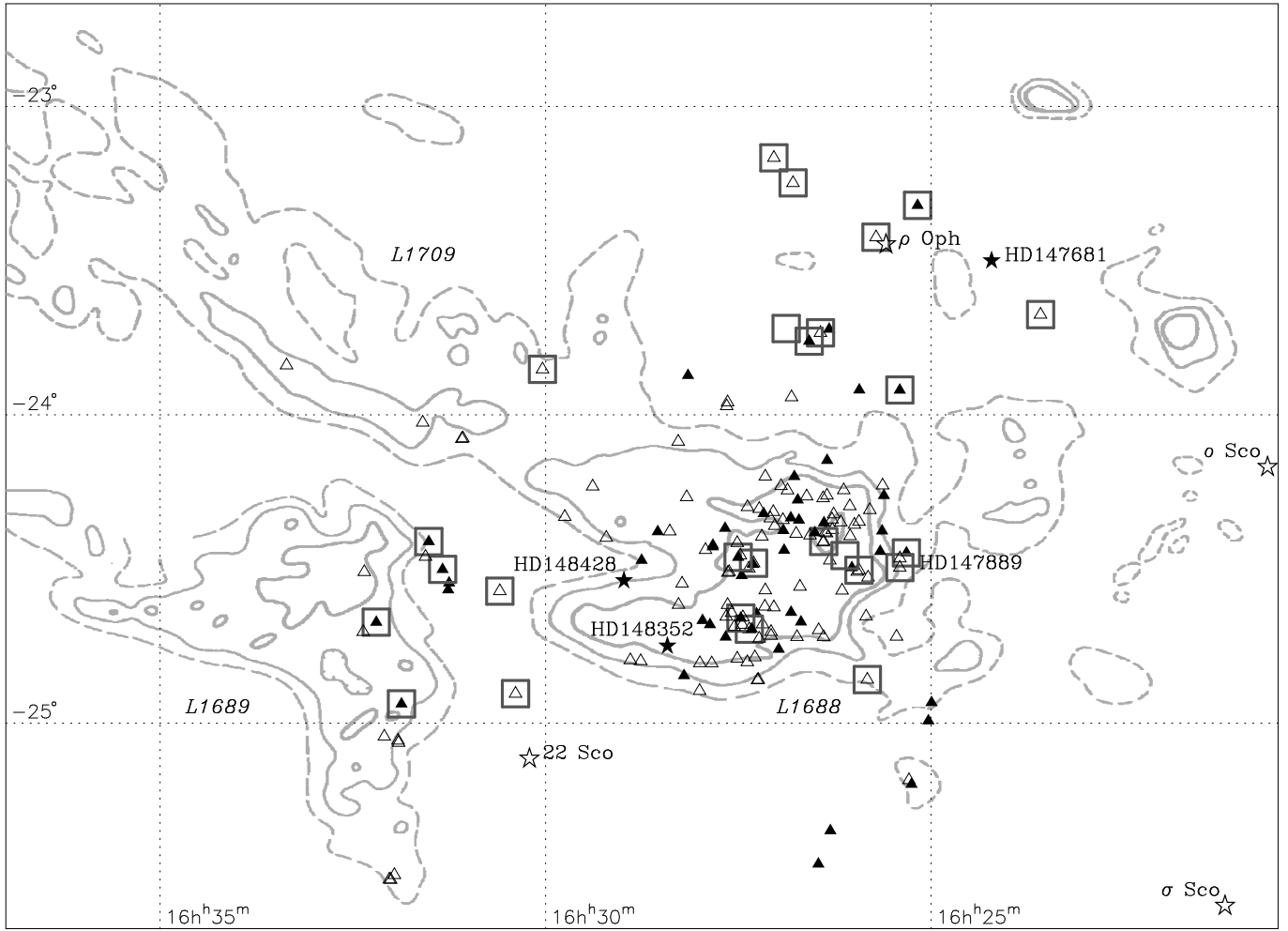}
\caption{The $^{13}\mathrm{CO\,} (J=0-1)$ contours of the $\rho$ Oph molecular clouds for $T^*_A(^{13}\mathrm{CO})= 2, 6, 10 \mathrm{\ and\ } 20\mathrm{\,K}$ \citep{loren1}. Each triangle marks a star of our sample. Filled triangles indicate double or multiple systems. The squares frame those areas that are used for the determination of the stellar background density. With exception of two cases the squares are centered around  or near stars included in our sample. While the empty square close to the center of the core contains \object{ISO-Oph~13} and \object{ISO-Oph~14}, the second one southeast of $\rho$ Oph is centered around \object{VSS~28}. Bright prominent stars not included in our sample are marked by a white asterisk, while black ones represent stars used as a PSF reference. All coordinates are in equinox J2000.0.}
\label{cloudcomplex}
\end{figure*}


\section{The cloud complex}

The $\rho$~Ophiuchi Dark Cloud (\object{L1688}, see Fig.\ref{cloudcomplex}) is the densest part of a complex of vast dark nebulae and molecular clouds that extends from $\mathrm{l} \approx 345^{\circ}$ to $10^{\circ}$ and from $\mathrm{b} \approx0^{\circ}$ to $+25^{\circ}$. The eastern part of this complex is dominated by long elongated filaments. A scenario presented by \cite{degeus2} assumes that early-type stars located in the Upper-Scorpius OB association ($\mathrm{l} \approx 360^{\circ} \dots 343^{\circ}, \mathrm{b} \approx +10^{\circ} \dots +30^{\circ}$) produced a shock-wave that encountered the dense precursor of the $\rho$~Oph cloud from behind, swept away material and deposited it in the present day filaments. This encounter may also have triggered the continuing low-mass star formation within this cloud, resulting in an extremely young population of stars with a median age of $\approx 0.3 \mathrm{\,Myr}$ \citep{greene, luhman}.

A recent paper by \cite{sartori} investigates the star-formation process on a larger scale. They found that the pre-main-sequence stars within the Ophiuchus, Lupus and Chamaeleon molecular cloud complexes follow a similar spatial distribution as the early-type stars in the subgroups of the Scorpius-Centaurus OB association and a newly found OB association in Chamaeleon. Furthermore, the young objects form an almost uniform group with respect to their kinematics and ages. The most natural scenario to explain the measurements is a spiral arm passing close to the Sun. The global distribution of HII regions \citep{lepine} supports this hypothesis.

\begin{table*}
\caption[]{Contributions from different papers for 
$m_\mathrm{K} = m_\mathrm{2MASS} < m_{\mathrm{lim}} = \mathrm{10.5\,mag}$}
\label{selection}
$$
\begin{array}{cp{0.23\linewidth}rp{0.015\linewidth}rrrrp{0.015\linewidth}p{0.16\linewidth}p{0.15\linewidth}p{0.10\linewidth}}
\hline
\noalign{\smallskip}
\# & Paper & \mathrm{sources} & & \multicolumn{3}{c}{m_K < m_{\mathrm{lim}}}   & \mathrm{{\bf survey}} & & region & criterion & association\\
   &       &                  & & \mathrm{total} & \mathrm{new} & \mathrm{obs} &                       & &        &           &\\
\noalign{\smallskip}
\hline
\noalign{\smallskip}
1  & \cite{casanova} (Table 2) &  87 & & 61 & 61 & 59     & {\bf 61} & & core						& X-ray + NIR, visual & bona fide\\
2  & \cite{casanova} (Table 1) &  19 & &  8 &  8 &  7     & {\bf 7}  & & core						& X-ray + NIR	      & probable\\
3  & \cite{casanova} (Table 3) &  22 & &  2 &  2 &  2     & {\bf 2}  & & core						& X-ray + NIR	      & candidate\\
4  & \cite{wly} (Table 4)      &  74 & & 56 &  5 &  5     & {\bf 5}  & & \object{L1688} 				& visual to FIR SED   & bona fide\\
5  & \cite{gwayl}              &  47 & & 37 & 14 & 14     & {\bf 14} & & \object{L1688}, \object{L1689}, \object{L1709} & visual to MIR SED   & bona fide\\
6  & \cite{ba}                 &  30 & & 30 & 13 & 13     & {\bf 13} & & whole complex  				& visual spectra      & bona fide\\
7  & \cite{grosso}             &  54 & & 46 &  9 &  7     & {\bf 7}  & & core						& Xray + NIR, MIR     & bona fide\\
8  & \cite{iso-oph}            & 212 & & 98 & 20 & 15     & {\bf 15} & & \object{L1688}, \object{L1689} 		& MIR excess	      & bona fide\\
9  & \cite{hbc}                &  24 & & 24 &  6 &  6     & {\bf 6}  & & whole complex  				& visual spectra      & bona fide\\
10 & \cite{wsb}                &  57 & & 53 & 19 & 15     & {\bf 15} & & whole complex  				& H$\alpha$	      & probable\\
11 & \cite{elias}              &  26 & & 26 &  3 &  3     & {\bf 3}  & & whole complex  				& NIR to MIR SED      & bona fide\\
12 & \cite{wly} (Table 6)      &  38 & & 24 & 13 & 10     & {\bf 10} & & \object{L1688} 				& visual to FIR SED   & candidate\\
\noalign{\smallskip}
 & & & & & \Sigma & & {\bf 158} & & \\
\noalign{\smallskip}
\hline
\end{array}
$$
\end{table*}

The distance to the $\rho$ Oph cloud is still the subject of discussions. A lower limit was published by \cite{knude} who found a steep increase of reddening at $120\mathrm{\,pc}$. An upper limit of $145\pm2\mathrm{\,pc}$ has been derived by \cite{zeeuw}. They determined the distance of the Scorpius-Centaurus OB association in the background of the $\rho$~Oph cloud by analysing positions, proper motions, and parallaxes of its members.  In this paper we assume a value of $140\mathrm{\,pc}$. The same distance as measured for the Taurus-Auriga association and used by \cite{koehler1}. This allows a direct comparison of the results.


\section{The Sample}

Our sample of 158 young stellar objects (YSOs) recruits from surveys at optical, infrared, and X-ray wavelengths (Table~\ref{selection}). From these surveys we selected objects which can be considered as cloud members using criteria commonly applied in distinguishing young stars from background or foreground stars. The most convincing are detailed studies of the optical spectra \citep{hbc, ba}, infrared spectral energy distributions \citep{wly, gwayl, elias}, mid-infrared colour-magnitude relations \citep{iso-oph}, and X-ray detections combined with optical/infrared information \citep{casanova, grosso}. We tried to combine and observe a sample as reliable and complete as possible down to magnitude $\mathrm{m}_\mathrm{K}\leq 10.5\mathrm{\ mag}$. Therefore, we preferably included objects fulfilling more than one of the criteria infrared excess, X-ray detection, and H$\alpha$ emission. The criteria met by the individual sources of our sample are indicated in Table~\ref{Sample} together with the number of the catalogue that lead to their selection. If the criteria did not appear strong, we marked the source with an `U'. Our sample may be characterised as volume-limited with magnitude cutoff. It was intended to be larger in size and more statistically complete than earlier surveys.

The coordinates and the magnitudes in the K-band presented in Table~\ref{Sample} are taken from the Two Micron All-Sky-Survey (2MASS) Catalog of Point Sources. At the time of the preparation of our survey, we had the slightly different K-band magnitudes of \cite{gy} and \cite{bklt} available and used them to determine the magnitude cutoff. This means that some sources close to the cutoff and bright enough in the 2MASS survey were not observed, and vice versa.

We started to build the sample with the then new list of 87\footnote{To allow easier comparison of different papers the wide binaries \object{SR~12}, \object{SR~24}, \object{ROXs~31}, \object{ROXs~43} are always counted as one object.} confirmed cloud members presented in \cite{casanova}. These authors  analysed a deep ROSAT image of the central region of the $\rho$~Oph star-forming region and compared the sources with a list of confirmed members mainly derived from the infrared surveys of \cite{wly} and \cite{gwayl}. This list was completed by including \object{ROXs~4}, \object{SR~2} and \object{VLA~1623}. It includes X-ray sources with an IR-counterpart but not detected in the visible. This strengthens the role of X-ray observations as a criterion of cloud membership. Since in \cite{casanova} 67\% of the found X-ray sources and 42\% of the candidate X-ray sources are common with the list of confirmed cloud members, hitherto unconfirmed cloud members coinciding with the remaining (candidate) X-ray sources are probable new cloud members and thus also targets of our survey. After removal of the background giant \object{VSSG~6} \citep{luhman} we are left with 61 certain and 10 probable cloud members (see Table~\ref{selection}). For the catalogue differences just mentioned \object{IRS~46} and \object{IRS~54}, bright enough in 2MASS, did not make it into our sample, while \object{WL~5} and \object{WL~6} were observed.

From the table of cluster members in \cite{wly}, \object{H$\alpha$~38}, \object{H$\alpha$~60}, \object{SR~20} and \object{H$\alpha$~63} are missing in the list of \cite{casanova} due to their position outside the investigated core region, as well as the objects \object{IRS~7}, \object{IRS~8}, \object{IRS~14} where the IRAS association was uncertain. Also the source \object{VSSG~12} was ignored for inconsistencies in the coordinates\footnote{Although SIMBAD identifies \object{VSSG~12} with \object{GSS~30-IRS~2} \cite{wly} give other coordinates north of \object{GSS~29}.}. With exception of the spurious \object{VSSG~12} and the faint objects \object{IRS~7} and \object{IRS~14} we reinserted these sources.

The multicolour infrared study by \cite{gwayl} includes also sources in \object{L1689} and \object{L1709}. We removed \object{VSSG~13}, \object{VSSG~15}, and \object{VSSG~16}, because they had been identified by \cite{elias} as field sources, and also the background giants \object{GY~45}, \object{GY~65},\object{GY~232}, \object{GY~411}, and \object{VSSG~6} \citep{luhman}. From the remaining 37 sources brighter than $\mathrm{m}_\mathrm{K}=10.5\mathrm{\,mag}$ 14 were new and are part of our sample.

\cite{ba} searched for counterpart candidates of X-ray sources detected with the Einstein satellite. Studying 46 optically visible stars lying in the error circles of 29 ROX sources with spectroscopic and photometric methods resulted in the identification of 29 certain and one probable (\object{ROXs 45D} = \object{DoAr~48}) cloud member. Of these, we added the 13 until now unaccounted cloud members to our list.

Out of 63 sources found with the ROSAT High Resolution Imager \cite{grosso} could identify 54 with optical, infrared and radio sources. This emphasises again the usefulness of X-ray emission as criterion for membership. We observed seven of the nine new targets, omitting two which were close to the brightness limit.

Recently \citep{iso-oph} presented an extensive mid-infrared survey of \object{L1688}, \object{L1689N} and \object{L1689S} performed with the ISOCAM camera on board the ISO-satellite at 6.7~$\mu$m and 14.3~$\mu$m. A catalogue of 212 sources detected at both wavelengths and classified as cloud members on the basis of colour-magnitude relations is now available. This catalogue includes 98 objects brighter than $\mathrm{m}_\mathrm{K}=10.5\mathrm{\,mag}$, of which twenty were not already included in our list. With exception of five sources close to the brightness limit all could be observed. If a source of our survey is included in this catalogue, the number therein is given in the second column of Table~\ref{Sample}.

In the third edition of their catalogue \cite{hbc} listed 24 sources towards the $\rho$ Oph molecular clouds. In our complete field six sources were new and thus added to our list. 

\cite{wsb} used objective-prism plates to survey 40 square degrees toward the Sco-Cen OB association including much of the $\rho$~Oph cloud complex for H$\alpha$ emisssion. Of the 57 objects not far from the central cloud \object{L1688}, nineteen were not yet included in our sample. They are mainly located in the western part of the complex. All sources of this catalogue that have been observed are indicated by their number in the third column of Table~\ref{Sample}. From the sources in \cite{elias} 26 fall into our region, of which three add to our catalogue. From the list of unidentified sources given in \cite{wly} thirteen sources are new and we could observe 10 of them.

The full list of sources brighter than 10.5~mag in the K-band would include 173 objects spread over the molecular clouds with a natural concentration in \object{L1688}. Our multiplicity survey covers 156 of these and 2 slightly fainter young stellar objects. Eleven well known sources among them have been already observed in the last decade by \cite{ghez} and \cite{sgl} with speckle imaging and during lunar occultations, i.e. with sufficient resolution and sensitivity. So, there was no necessity to observe these sources again. They are marked in Table~\ref{Sample} with an `O'.

\begin{table*}
\caption[]{The 158 sources of our survey}
\label{Sample}
$$
\begin{array}{p{0.14\linewidth}ccp{0.086\linewidth}p{0.085\linewidth}p{0.086\linewidth}p{0.085\linewidth}cccp{0.034\linewidth}p{0.030\linewidth}p{0.15\linewidth}}
\hline
\noalign{\smallskip}
Object & \mathrm{ISO} & \mathrm{H}\alpha & \multicolumn{2}{c}{J2000.0} & \multicolumn{2}{c}{B1950.0} & \mathrm{K}_{\mathrm{2MASS}} & \mathrm{K}_{\mathrm{BKLT}} & \multicolumn{2}{c}{\mathrm{Selection}} & Note & Other Designation\\
 & \mathrm{No.} & \mathrm{No.} &  \hspace{0.7cm}$\alpha$ & \hspace{0.6cm}$\delta$ & \hspace{0.7cm}$\alpha$ & \hspace{0.6cm}$\delta$ & \mathrm{[mag]} & \mathrm{[mag]} & \mathrm{\#} & \hspace{0.03cm}crit. &  & \\
\noalign{\smallskip}
\hline
\noalign{\smallskip}
\object{H$\alpha$ 16}    &   - &    16 & 16 23 34.83& -23 40 30.3  & 16 20 34.46  & -23 33 35.0	&  8.552 &     - & 10 & -- v  -- & U  & -\\
\object{H$\alpha$ 18}    &   - &    18 & 16 24 59.74& -24 56 00.8  & 16 21 57.68  & -24 49 11.1	&  9.441 &     - & 10 & -- v  -- & -  & -\\
\object{H$\alpha$ 19}    &   - &    19 & 16 25 02.09& -24 59 32.3  & 16 21 59.95  & -24 52 42.8	&  9.220 &     - & 10 & -- v  -- & -  & -\\
\object{Haro 1-4}        &   - &    20 & 16 25 10.52& -23 19 14.6  & 16 22 10.51  & -23 12 25.7	&  7.509 &     - &  9 & -- v  -- & O  & \object{HBC~257}\\
\object{H$\alpha$ 21}    &   - &    21 & 16 25 15.22& -25 11 54.1  & 16 22 12.80  & -25 05 05.4	&  9.642 &     - & 10 & -- v  -- & U  & -\\
\object{H$\alpha$ 22}    &   - &    22 & 16 25 17.27& -25 11 05.5  & 16 22 14.87  & -25 04 16.9	&  9.911 &     - & 10 & -- v  -- & -  & -\\
\object{SR 2}            &   - &     - & 16 25 19.24& -24 26 52.6  & 16 22 17.79  & -24 20 04.2	&  7.798 &     - &  1 & x  -- i  & A  & \object{Elias~6}\\
\object{SR 22}           &   - &    23 & 16 25 24.34& -24 29 44.3  & 16 22 22.83  & -24 22 56.2	&  9.446 &     - &  1 & x  v  i  & -  & \object{V 852~Oph}, \object{HBC~258}\\
\object{SR 1}            &   - &     - & 16 25 24.31& -24 27 56.6  & 16 22 22.83  & -24 21 08.5	&  4.582 &     - &  1 & x  -- i  & AE & \object{Elias~9}\\
\object{ROXs 2}          &   - &     - & 16 25 24.35& -23 55 10.3  & 16 22 23.57  & -23 48 22.3	&  8.379 &     - &  6 & x  v  -- & B  & \object{HBC~635}\\
\object{SR 8}            &   - &     - & 16 25 26.87& -24 43 09.0  & 16 22 25.06  & -24 36 21.1	&  8.662 &  8.72 &  2 & x  -- -- & -  & -\\
\object{IRS 2}           &   1 &     - & 16 25 36.74& -24 15 42.4  & 16 22 35.51  & -24 08 55.2	&  8.380 &  8.36 &  1 & x  -- i  & -  & -\\
\object{Elias 12}        &   - &     - & 16 25 37.81& -24 13 43.6  & 16 22 36.62  & -24 06 56.5	&  6.867 &  6.95 & 11 & -- -- i  & -  & -\\
\object{J162538-242238}  &   2 &     - & 16 25 38.12& -24 22 36.3  & 16 22 36.75  & -24 15 49.1	&  9.545 &  9.68 &  7 & x  -- i  & -  & -\\
\object{IRS 3}           &   3 &     - & 16 25 39.58& -24 26 34.9  & 16 22 38.12  & -24 19 47.9	&  8.954 &  8.93 &  2 & x  -- i  & -  & -\\
\object{H$\alpha$ 24}    &   - &    24 & 16 25 42.89& -23 25 26.1  & 16 22 42.73  & -23 18 39.3	&  9.204 &     - & 10 & -- v  -- & -  & -\\
\object{IRS 8}           &   - &     - & 16 25 47.69& -24 18 31.1  & 16 22 46.39  & -24 11 44.6	&  9.701 &  9.69 &  4 & -- -- i  & -  & -\\
\object{IRS 9}           &   - &     - & 16 25 49.05& -24 31 39.1  & 16 22 47.48  & -24 24 52.7	&  9.487 &  9.41 &  1 & x  -- i  & -  & -\\
\object{ROXs 3}          &   - &     - & 16 25 49.64& -24 51 31.9  & 16 22 47.64  & -24 44 45.5	&  8.784 &  8.78 &  6 & x  v  -- & -  & \object{HBC~636}\\
\object{VSS 23}          &   5 &     - & 16 25 50.53& -24 39 14.5  & 16 22 48.78  & -24 32 28.2	&  8.331 &  8.24 &  1 & x  v  i  & -  & \object{IRS~10}, \object{ROXs~4}\\
\object{ROXs 5}          &   - &     - & 16 25 55.83& -23 55 10.0  & 16 22 55.02  & -23 48 24.0	&  8.383 &     - &  6 & x  v  -- & B  & -\\
\object{IRS 11}          &   - &     - & 16 25 56.10& -24 30 14.9  & 16 22 54.54  & -24 23 28.9	&  9.764 &  9.76 & 12 & -- -- i  & U  & -\\
\object{SR 4}            &   6 &    25 & 16 25 56.16& -24 20 48.2  & 16 22 54.81  & -24 14 02.3	&  7.518 &  7.25 &  1 & x  v  i  & O  & \object{ROXs~6}, \object{HBC~259}, \object{Elias~13}\\
\object{GSS 20}          &   7 &     - & 16 25 57.52& -24 30 31.7  & 16 22 55.96  & -24 23 45.9	&  8.319 &  8.37 &  1 & x  v  i  & -  & \object{ROXs~7}\\
\object{Chini 8}         &   - &     - & 16 25 59.65& -24 21 22.3  & 16 22 58.28  & -24 14 36.6	&  9.531 &  9.52 &  2 & x  -- -- & -  & -\\
\object{ROXR1-12}        &   8 &     - & 16 26 01.61& -24 29 45.0  & 16 23 00.06  & -24 22 59.4	&  9.629 &  9.55 &  7 & x  -- i  & -  & \object{J162601-242945}\\
\object{DoAr 21}         &  10 &     - & 16 26 03.02& -24 23 36.0  & 16 23 01.60  & -24 16 50.6	&  6.227 &  6.16 &  1 & x  v  i  & O  & \object{ROXs~8}, \object{HBC~637}, \object{Elias~14}\\
\object{VSSG 19}         &  11 &     - & 16 26 03.29& -24 17 46.5  & 16 23 02.00  & -24 11 01.0	&  9.115 &  9.07 &  5 & x  -- i  & -  & -\\
\object{Chini 11}        &   - &     - & 16 26 08.01& -24 14 39.2  & 16 23 06.77  & -24 07 54.1	&  9.812 &  9.79 & 12 & -- -- i  & U  & -\\
\object{SR 3}            &  16 &     - & 16 26 09.31& -24 34 12.1  & 16 23 07.66  & -24 27 27.1	&  6.504 &  6.42 &  1 & x  -- i  & A  & \object{Elias~16}\\
\object{GSS 26}          &  17 &     - & 16 26 10.33& -24 20 54.8  & 16 23 08.96  & -24 14 09.8	&  8.475 &  9.38 &  1 & x  -- i  & -  & -\\
\object{SKS 1-7}         &  18 &     - & 16 26 15.81& -24 19 22.1  & 16 23 14.47  & -24 12 37.5	& 10.031 &  9.98 &  8 & -- -- i  & -  & -\\
\object{GSS 29}          &  19 &     - & 16 26 16.84& -24 22 23.2  & 16 23 15.44  & -24 15 38.6	&  8.201 &  8.19 &  1 & x  -- i  & -  & \object{Elias~18}\\
\object{DoAr 24}         &  20 &    27 & 16 26 17.06& -24 20 21.6  & 16 23 15.70  & -24 13 37.1	&  8.063 &  8.09 &  1 & x  v  i  & O  & \object{GSS~28}, \object{ROXs~10A}, \object{HBC~638}, \object{Elias~19}\\
\object{H$\alpha$ 26}    &   - &    26 & 16 26 18.40& -25 20 55.7  & 16 23 15.73  & -25 14 11.2	&  8.934 &     - & 10 & -- v  -- & -  & \object{DoAr23}\\
\object{VSSG 1}          &  24 &     - & 16 26 18.87& -24 28 19.7  & 16 23 17.33  & -24 21 35.3	&  8.072 &  8.68 &  1 & x  -- i  & -  & \object{Elias~20}\\
\object{DoAr 22}         &   - &     - & 16 26 19.32& -23 43 20.5  & 16 23 18.75  & -23 36 36.2	&  9.018 &     - &  9 & -- v  -- & -  & \object{HBC~260}\\
\object{H$\alpha$ 28}    &  27 &    28 & 16 26 20.97& -24 08 51.9  & 16 23 19.85  & -24 02 07.6	&  9.502 &  9.48 &  7 & x  v  i  & C  & -\\
\object{J162621-241544}  &  28 &     - & 16 26 21.02& -24 15 41.5  & 16 23 19.75  & -24 08 57.2	&  9.273 &  9.27 &  8 & -- -- i  & -  & -\\
\object{Elias 21}        &  29 &     - & 16 26 21.38& -24 23 04.1  & 16 23 19.96  & -24 16 19.8	&  8.835 &  8.32 &  1 & x  -- i  & CE & \object{GSS 30 - IRS 1}\\
\object{GSS 30 - IRS 2}  &  34 &     - & 16 26 22.39& -24 22 52.9  & 16 23 20.96  & -24 16 08.8	&  9.957 &  9.60 &  1 & x  -- i  & -  & -\\
\object{DoAr 24E}        &  36 &    30 & 16 26 23.36& -24 20 59.8  & 16 23 21.98  & -24 14 15.7	&  6.571 &  6.44 &  1 & x  v  i  & O  & \object{GSS 31}, \object{ROXs~10B}, \object{HBC~639}, \object{Elias~22}\\
\object{LFAM 3}          &  37 &     - & 16 26 23.58& -24 24 39.5  & 16 23 22.12  & -24 17 55.4	& 10.224 &  9.94 &  1 & x  -- i  & -  & \object{GY 21}\\
\object{DoAr 25}         &  38 &    29 & 16 26 23.68& -24 43 13.9  & 16 23 21.82  & -24 36 29.8	&  7.847 &  7.57 &  1 & x  v  i  & A  & -\\
\object{GSS 32}          &  39 &     - & 16 26 24.04& -24 24 48.1  & 16 23 22.58  & -24 18 04.0	&  7.324 &  7.20 &  1 & x  -- i  & C  & \object{S~2}, \object{Elias~23}, \object{GY 23}\\
\object{Elias 24}        &  40 &    31 & 16 26 24.07& -24 16 13.5  & 16 23 22.79  & -24 09 29.4	&  6.685 &  6.77 &  1 & x  v  i  & -  & -\\
\object{H$\alpha$ 33}    &   - &    33 & 16 26 26.06& -23 44 02.6  & 16 23 25.46  & -23 37 18.7	&  9.928 &     - & 10 & -- v  -- & U  & -\\
\object{GY 33}           &  43 &     - & 16 26 27.54& -24 41 53.5  & 16 23 25.70  & -24 35 09.7	&  9.983 &  9.83 &  8 & -- -- i  & -  & -\\
\object{ROXs 12}         &   - &    32 & 16 26 27.75& -25 27 24.7  & 16 23 24.93  & -25 20 40.8	&  9.211 &     - &  6 & x  v  -- & -  & -\\
\object{VSSG 27}         &  46 &     - & 16 26 30.47& -24 22 57.1  & 16 23 29.04  & -24 16 13.5	&  9.977 & 10.72 &  1 & x  -- i  & -  & -\\
\object{S1}              &  48 &     - & 16 26 34.17& -24 23 28.3  & 16 23 32.72  & -24 16 44.9	&  6.317 &  6.32 &  1 & x  v  i  & -  & \object{ROXs~14}, \object{Elias~25}\\
\object{H$\alpha$ 35}    &   - &    35 & 16 26 34.90& -23 45 40.6  & 16 23 34.26  & -23 38 57.3	&  8.834 &     - & 10 & -- v  -- & -  & -\\
\noalign{\smallskip} 														  
\hline 																  
\end{array}																  
$$																	  
\end{table*}

\setcounter{table}{1}
\begin{table*}
\caption[]{(continued)}
$$
\begin{array}{p{0.14\linewidth}ccp{0.086\linewidth}p{0.085\linewidth}p{0.086\linewidth}p{0.085\linewidth}cccp{0.034\linewidth}p{0.030\linewidth}p{0.15\linewidth}}
\hline
\noalign{\smallskip}
Object & \mathrm{ISO} & \mathrm{H}\alpha & \multicolumn{2}{c}{J2000.0} & \multicolumn{2}{c}{B1950.0} & \mathrm{K}_{\mathrm{2MASS}} & \mathrm{K}_{\mathrm{BKLT}} & \multicolumn{2}{c}{\mathrm{Selection}} & Note & Other Designation\\
 & \mathrm{No.} & \mathrm{No.} &  \hspace{0.7cm}$\alpha$ & \hspace{0.6cm}$\delta$ & \hspace{0.7cm}$\alpha$ & \hspace{0.6cm}$\delta$ & \mathrm{[mag]} & \mathrm{[mag]} & \mathrm{\#} & \hspace{0.03cm}crit. &  & \\
\noalign{\smallskip}
\hline
\noalign{\smallskip}
\object{J162636-241554}  &  51 &     - & 16 26 36.83  & -24 15 51.9     & 16 23 35.54    & -24 09 08.7     &  9.589   &  9.38 &  7 & x  -- i  & -  & -\\
\object{H$\alpha$ 37}    &  56 &    37 & 16 26 41.26  & -24 40 18.0	& 16 23 39.44    & -24 33 35.0     &  9.273   &  9.20 &  1 & x  v  i  & -  & -\\
\object{WL 8}            &  58 &     - & 16 26 42.02  & -24 33 26.2	& 16 23 40.35    & -24 26 43.4     &  9.578   &  9.44 &  8 & -- -- i  & -  & -\\
\object{GSS 37}          &  62 &     - & 16 26 42.86  & -24 20 29.9	& 16 23 41.47    & -24 13 47.1     &  7.878   &  8.00 &  1 & x  -- i  & -  & \object{VSSG~2}, \object{Elias~26}\\
\object{VSSG 11}         &  64 &     - & 16 26 43.76  & -24 16 33.3	& 16 23 42.45    & -24 09 50.6     &  9.604   &  9.58 &  1 & x  -- i  & -  & -\\
\object{GY 112}          &  66 &     - & 16 26 44.30  & -24 43 14.1	& 16 23 42.42    & -24 36 31.4     &  9.573   &  9.54 &  2 & x  -- i  & -  & -\\
\object{GSS 39}          &  67 &     - & 16 26 45.03  & -24 23 07.7	& 16 23 43.58    & -24 16 25.1     &  8.955   &  8.88 &  1 & x  -- i  & -  & \object{Elias~27}\\
\object{ROXs 16}         &  68 &    38 & 16 26 46.43  & -24 12 00.1	& 16 23 45.22    & -24 05 17.5     &  7.485   &  7.51 &  4 & x  v  i  & -  & \object{VSS~27}\\
\object{Haro 1-8}        &   - &    39 & 16 26 47.42  & -23 14 52.2	& 16 23 47.42    & -23 08 09.7     &  8.619   &     - &  9 & -- v  -- & B  & \object{HBC~261}\\
\object{H$\alpha$ 40}    &   - &    40 & 16 26 48.65  & -23 56 34.2	& 16 23 47.76    & -23 49 51.8     &  8.449   &     - & 10 & -- v  -- & -  & -\\
\object{WL 18}           &  72 &     - & 16 26 48.98  & -24 38 25.2	& 16 23 47.20    & -24 31 42.8     &  9.977   &  9.82 &  1 & x  -- i  & E  & -\\
\object{VSSG 3}          &  73 &     - & 16 26 49.23  & -24 20 02.9	& 16 23 47.85    & -24 13 20.5     &  8.687   &  8.62 &  1 & x  -- i  & -  & -\\
\object{VSSG 10}         &   - &     - & 16 26 51.69  & -24 14 41.6	& 16 23 50.41    & -24 07 59.3     &  9.713   &  9.77 & 12 & -- -- i  & U  & -\\
\object{VSSG 5}          &  78 &     - & 16 26 54.44  & -24 26 20.7	& 16 23 52.92    & -24 19 38.7     & 10.014   &  9.88 &  1 & x  -- i  & -  & -\\
\object{GY 156}          &  80 &     - & 16 26 54.97  & -24 22 29.7	& 16 23 53.53    & -24 15 47.7     & 10.163   & 10.19 &  3 & x  -- i  & -  & -\\
\object{VSSG 7}          &  81 &     - & 16 26 55.31  & -24 20 27.8	& 16 23 53.91    & -24 13 45.8     &  9.789   &  9.69 &  8 & -- -- i  & -  & -\\
\object{J162656-241353}  &  83 &     - & 16 26 56.77  & -24 13 51.6	& 16 23 55.51    & -24 07 09.7     &  9.251   &  9.57 &  8 & -- -- i  & -  & -\\
\object{SR 24}           &  88 & 41/42 & 16 26 58.51  & -24 45 36.9	& 16 23 56.56    & -24 38 55.1     &  7.057   &  7.08 &  1 & x  v  i  & CO & \object{HBC~262}, \object{Elias~28}\\
\object{VSSG 8}          &  91 &     - & 16 27 01.62  & -24 21 37.0	& 16 24 00.19    & -24 14 55.5     &  9.389   &  9.32 &  8 & -- -- i  & -  & -\\
\object{H$\alpha$ 44}    &   - &    44 & 16 27 02.37  & -23 09 59.2	& 16 24 02.46    & -23 03 17.8     &  9.435   &     - & 10 & -- v  -- & U  & -\\
\object{WL 16}           &  92 &     - & 16 27 02.34  & -24 37 27.2	& 16 24 00.57    & -24 30 45.7     &  8.064   &  7.92 &  1 & x  -- i  & -  & -\\
\object{VSSG 9}          &   - &     - & 16 27 02.85  & -24 18 54.7	& 16 24 01.48    & -24 12 13.2     & 10.116   & 10.07 & 12 & -- -- i  & U  & -\\
\object{GY 193}          &  96 &     - & 16 27 04.52  & -24 42 59.7	& 16 24 02.62    & -24 36 18.3     &  9.837   &  9.80 &  2 & x  -- i  & -  & -\\
\object{GY 194}          &  97 &     - & 16 27 04.56  & -24 42 14.0	& 16 24 02.69    & -24 35 32.6     &  9.809   &  9.82 &  2 & x  -- i  & -  & -\\
\object{VSSG 21}         &   - &     - & 16 27 05.16  & -24 20 07.7	& 16 24 03.76    & -24 13 26.4     &  9.374   &  9.27 & 12 & -- -- i  & U  & -\\
\object{J162708-241204}  & 106 &     - & 16 27 09.07  & -24 12 00.8	& 16 24 07.84    & -24 05 19.7     &  9.800   &  9.76 &  8 & -- -- i  & -  & -\\
\object{WL 10}           & 105 &     - & 16 27 09.10  & -24 34 08.1	& 16 24 07.40    & -24 27 27.1     &  8.915   &  8.85 &  1 & x  -- i  & -  & -\\
\object{Elias 29}        & 108 &     - & 16 27 09.43  & -24 37 18.8	& 16 24 07.66    & -24 30 37.7     &  7.140   &  7.54 &  1 & x  -- i  & -  & -\\
\object{Elias 30}        & 110 &     - & 16 27 10.28  & -24 19 12.7	& 16 24 08.89    & -24 12 31.8     &  6.719   &  6.30 &  1 & x  -- i  & C  & \object{SR~21}\\
\object{GY 224}          & 112 &     - & 16 27 11.18  & -24 40 46.7	& 16 24 09.33    & -24 34 05.7     & 10.196   & 10.79 &  1 & x  -- i  & -  & -\\
\object{IRS 32}          & 113 &     - & 16 27 11.68  & -24 23 42.0	& 16 24 10.20    & -24 17 01.1     & 10.107   & 10.06 &  5 & -- -- i  & -  & -\\
\object{VSSG 24}         & 116 &     - & 16 27 13.73  & -24 18 16.9	& 16 24 12.36    & -24 11 36.1     &  9.287   &  9.32 &  8 & -- -- i  & -  & -\\
\object{IRS 32b}         & 117 &     - & 16 27 13.82  & -24 43 31.7	& 16 24 11.91    & -24 36 50.9     &   9.978  & 10.13 &  8 & -- -- i  & O  & -\\
\object{ROXs 20A}        &   - &    45 & 16 27 14.49  & -24 51 33.5	& 16 24 12.40    & -24 44 52.8     & 10.381   & 10.39 &  5 & x  v  i  & -  & \object{HBC~640}\\
\object{ROXs 20B}        &   - &    46 & 16 27 15.13  & -24 51 38.8	& 16 24 13.05    & -24 44 58.1     &  9.392   &  9.51 &  5 & x  v  i  & -  & \object{HBC~641}\\
\object{WL 20}           & 121 &     - & 16 27 15.88  & -24 38 43.4	& 16 24 14.07    & -24 32 02.8     &  9.590   &  9.21 &  1 & x  -- i  & CE & -\\
\object{H$\alpha$ 47}    &   - &    47 & 16 27 17.08  & -24 47 11.2	& 16 24 15.08    & -24 40 30.6     &  9.487   &  9.47 & 10 & -- v  -- & -  & -\\
\object{WL 5}            & 125 &     - & 16 27 18.17  & -24 28 52.7	& 16 24 16.57    & -24 22 12.2     & 10.558   & 10.28 &  1 & x  -- i  & L  & -\\
\object{WL 4}            & 128 &     - & 16 27 18.49  & -24 29 05.9	& 16 24 16.88    & -24 22 25.5     &  9.683   &  9.13 &  1 & x  -- i  & -  & -\\
\object{SR 12}           & 130 &     - & 16 27 19.51  & -24 41 40.4	& 16 24 17.64    & -24 35 00.0     &   8.408  &  8.41 &  1 & x  v  i  & O  & \object{ROXs~21}, \object{HBC~263}\\
\object{IRS 42}          & 132 &     - & 16 27 21.47  & -24 41 43.1	& 16 24 19.59    & -24 35 02.8     &  8.483   &  8.41 &  1 & x  -- i  & -  & -\\
\object{WL 6}            & 134 &     - & 16 27 21.80  & -24 29 53.4	& 16 24 20.18    & -24 23 13.1     & 10.827   & 10.04 &  1 & x  -- i  & L  & -\\
\object{VSSG 22}         & 135 &     - & 16 27 22.91  & -24 17 57.4	& 16 24 21.54    & -24 11 17.2     &  9.454   &  9.41 &  1 & x  -- i  & -  & -\\
\object{H$\alpha$ 49}    &   - &    49 & 16 27 22.97  & -24 48 07.1	& 16 24 20.95    & -24 41 27.0     &  9.390   &  9.30 &  1 & x  v  i  & -  & -\\
\object{GY 262}          & 140 &     - & 16 27 26.49  & -24 39 23.1	& 16 24 24.66    & -24 32 43.2     &  9.952   &  9.77 &  1 & x  -- i  & -  & -\\
\object{IRS 43}          & 141 &     - & 16 27 26.94  & -24 40 50.8	& 16 24 25.07    & -24 34 10.9     &  9.745   &  9.46 &  1 & x  -- i  & -  & \object{YLW 15A}\\
\object{VSSG 25}         & 142 &     - & 16 27 27.38  & -24 31 16.6	& 16 24 25.72    & -24 24 36.7     &  9.316   &  9.30 &  1 & x  -- i  & -  & \object{Elias~31}\\
\object{IRS 44}          & 143 &     - & 16 27 28.03  & -24 39 33.5	& 16 24 26.19    & -24 32 53.7     & 10.379   &  9.65 &  1 & x  -- i  & -  & \object{YLW 16A}\\
\object{VSSG 18}         & 144 &     - & 16 27 28.45  & -24 27 21.0	& 16 24 26.87    & -24 20 41.3     & 10.101   &  9.39 &  1 & x  -- i  & C  & \object{Elias~32}\\
\object{VSSG 17}         & 147 &     - & 16 27 30.18  & -24 27 43.4	& 16 24 28.59    & -24 21 03.7     &  9.024   &  8.95 &  1 & x  -- i  & C  & \object{Elias~33}\\
\object{GY 284}          & 151 &     - & 16 27 30.84  & -24 24 56.0	& 16 24 29.32    & -24 18 16.4     & 10.070   & 10.04 &  3 & x  -- i  & -  & -\\
\object{J162730-244726}  & 149 &     - & 16 27 30.84  & -24 47 26.8	& 16 24 28.83    & -24 40 47.2     &  9.502   &  9.44 &  7 & x  -- i  & -  & -\\
\object{GY 292}          & 155 &     - & 16 27 33.11  & -24 41 15.3	& 16 24 31.23    & -24 34 35.8     &  7.806   &  7.92 &  1 & x  -- i  & -  & -\\
\object{H$\alpha$ 50}    & 156 &    50 & 16 27 35.26  & -24 38 33.4	& 16 24 33.44    & -24 31 54.1     &  9.668   &  9.64 &  5 & -- v  i  & -  & \object{GY~295}\\
\object{IRS 48}          & 159 &     - & 16 27 37.19  & -24 30 35.0	& 16 24 35.54    & -24 23 55.8     &  7.582   &  7.42 &  1 & x  -- i  & -  & -\\
\object{IRS 50}          &   - &     - & 16 27 38.13  & -24 30 42.9	& 16 24 36.47    & -24 24 03.8     &  9.658   &  9.59 &  1 & x  -- i  & -  & -\\
\noalign{\smallskip} 															     
\hline 																	     
\end{array}
$$
\end{table*}

\setcounter{table}{1}
\begin{table*}
\caption[]{(continued)}
$$
\begin{array}{p{0.14\linewidth}ccp{0.086\linewidth}p{0.085\linewidth}p{0.086\linewidth}p{0.085\linewidth}cccp{0.034\linewidth}p{0.030\linewidth}p{0.15\linewidth}}
\hline
\noalign{\smallskip}
Object & \mathrm{ISO} & \mathrm{H}\alpha & \multicolumn{2}{c}{J2000.0} & \multicolumn{2}{c}{B1950.0} & \mathrm{K}_{\mathrm{2MASS}} & \mathrm{K}_{\mathrm{BKLT}} & \multicolumn{2}{c}{\mathrm{Selection}} & Note & Other Designations\\
 & \mathrm{No.} & \mathrm{No.} &  \hspace{0.7cm}$\alpha$ & \hspace{0.6cm}$\delta$ & \hspace{0.7cm}$\alpha$ & \hspace{0.6cm}$\delta$ & \mathrm{[mag]} & \mathrm{[mag]} & \mathrm{\#} & \hspace{0.03cm}crit. &  & \\
\noalign{\smallskip}
\hline
\noalign{\smallskip}
\object{IRS 49}          & 163 &     - & 16 27 38.32  & -24 36 58.6	& 16 24 36.53    & -24 30 19.5     &  8.271   &  8.31 &  1 & x  -- i  & -  & -\\
\object{ROXs 30B}        &   - &    51 & 16 27 38.33  & -23 57 32.4	& 16 24 37.38    & -23 50 53.3     &  7.940   &     - &  6 & x  v  -- & -  & \object{DoAr~32}\\
\object{ROXs 30C}        &   - &    53 & 16 27 39.01  & -23 58 18.7	& 16 24 38.05    & -23 51 39.7     &  8.206   &     - &  6 & x  v  -- & -  & -\\
\object{H$\alpha$ 52}    & 166 &    52 & 16 27 39.43  & -24 39 15.5	& 16 24 37.59    & -24 32 36.5     &  8.464   &  8.35 &  1 & x  v  i  & -  & \object{GY~314}\\
\object{IRS 51}          & 167 &     - & 16 27 39.83  & -24 43 15.1	& 16 24 37.90    & -24 36 36.0     &  8.991   &  8.93 &  1 & x  -- i  & -  & -\\
\object{SR 9}            & 168 &    54 & 16 27 40.29  & -24 22 04.0	& 16 24 38.81    & -24 15 25.0     &  7.207   &  7.20 &  1 & x  v  i  & -  & \object{ROXs~29}, \object{HBC~264}, \object{Elias~34}\\
\object{GY 371}          & 178 &     - & 16 27 49.78  & -24 25 22.0	& 16 24 48.23    & -24 18 43.6     & 10.161   & 10.17 &  8 & -- -- i  & -  & -\\
\object{VSSG 14}         & 180 &     - & 16 27 49.87  & -24 25 40.2	& 16 24 48.31    & -24 19 01.9     &  7.301   &  7.32 &  1 & x  -- i  & -  & \object{Elias~36}\\
\object{IRS 56}          &   - &     - & 16 27 50.74  & -24 48 21.6	& 16 24 48.69    & -24 41 43.3     &  8.337   &  8.23 & 12 & -- -- i  & U  & -\\
\object{ROXs 31}         & 184 &     - & 16 27 52.09  & -24 40 50.4	& 16 24 50.20    & -24 34 12.2     &  8.126   &  8.09 &  1 & x  v  i  & -  & \object{IRS~55}, \object{HBC~642}\\
\object{SR 10}           & 187 &    57 & 16 27 55.58  & -24 26 17.9	& 16 24 54.00    & -24 19 39.9     &  8.896   &  8.74 &  1 & x  v  i  & -  & \object{HBC~265}\\
\object{GY 410}          & 188 &     - & 16 27 57.83  & -24 40 01.8	& 16 24 55.95    & -24 33 24.0     &  9.866   &  9.78 &  1 & x  -- i  & -  & -\\
\object{H$\alpha$ 58}    &   - &    58 & 16 27 59.97  & -24 48 19.3	& 16 24 57.91    & -24 41 41.6     &  9.269   &  9.26 &  2 & x  v  -- & U  & -\\
\object{J162800-245340}  &   - &     - & 16 28 00.11  & -24 53 42.7	& 16 24 57.94    & -24 47 05.1     &  9.651   &  9.63 &  7 & x  -- -- & U  & -\\
\object{H$\alpha$ 59}    &   - &    59 & 16 28 09.21  & -23 52 20.5	& 16 25 08.35    & -23 45 43.5     &  9.075   &     - & 10 & -- v  -- & U  & -\\
\object{VSS 35}          &   - &     - & 16 28 10.22  & -24 16 01.0	& 16 25 08.85    & -24 09 24.0     &  7.965   &  7.89 & 12 & -- -- i  & U  & -\\
\object{J162812-245043}  &   - &     - & 16 28 12.28  & -24 50 44.6	& 16 25 10.16    & -24 44 07.7     &  9.555   &  9.36 &  7 & x  -- -- & CU & -\\
\object{J162813-243249}  & 194 &     - & 16 28 13.79  & -24 32 49.4	& 16 25 12.06    & -24 26 12.7     & 10.096   & 10.04 &  8 & -- -- i  & -  & -\\
\object{H$\alpha$ 60}    & 196 &    60 & 16 28 16.51  & -24 36 58.0	& 16 25 14.68    & -24 30 21.4     &  9.316   &  9.43 &  4 & -- v  i  & -  & -\\
\object{ISO-Oph 195}     & 195 &     - & 16 28 16.73  & -24 05 14.3	& 16 25 15.59    & -23 58 37.7     &  8.860   &   -   &  8 & -- -- i  & -  & -\\
\object{SR 20 W (GWAYL)} &   - &     - & 16 28 23.33  & -24 22 40.6	& 16 25 21.81    & -24 16 04.5     &  8.623   &  8.55 &  5 & -- -- i  & -  & -\\
\object{SR 20}           & 198 &    61 & 16 28 32.66  & -24 22 44.9	& 16 25 31.13    & -24 16 09.4     &  6.850   &  7.16 &  4 & x  v  i  & O  & \object{ROXs~33}, \object{HBC~643}\\
\object{V 853 Oph}       & 199 &    62 & 16 28 45.28  & -24 28 19.0	& 16 25 43.61    & -24 21 44.4     &  7.997   &  7.88 &  6 & x  v  i  & O  & \object{ROXs~34}, \object{HBC~266}\\
\object{VSS 38}          &   - &     - & 16 28 45.98  & -24 47 55.3	& 16 25 43.89    & -24 41 20.7     &  5.960   &  6.14 & 12 & -- -- i  & U  & -\\
\object{H$\alpha$ 63}    &   - &    63 & 16 28 54.07  & -24 47 44.2	& 16 25 51.98    & -24 41 10.2     &  8.905   &  8.96 &  4 & -- v  i  & -  & -\\
\object{VSS 42}          &   - &     - & 16 29 12.73  & -24 23 55.3	& 16 26 11.14    & -24 17 22.6     &  5.825   &  6.27 & 12 & -- -- i  & U  & -\\
\object{IRAS 64a}        &   - &     - & 16 29 23.39  & -24 13 56.9	& 16 26 22.00    & -24 07 24.8     &  7.020   &  6.45 &  5 & -- -- i  & -  & -\\
\object{VSS 41}          &   - &     - & 16 29 45.12  & -24 19 50.5	& 16 26 43.59    & -24 13 19.9     &  8.664   &     - & 12 & -- -- i  & U  & -\\
\object{Elias 41}        &   - &     - & 16 30 02.41  & -23 51 09.1	& 16 27 01.47    & -23 44 39.7     &  6.905   &     - & 11 & -- -- i  & -  & -\\
\object{H$\alpha$ 67}    &   - &    67 & 16 30 23.40  & -24 54 16.2	& 16 27 21.08    & -24 47 48.1     &  9.293   &     - & 10 & -- v  -- & -  & -\\
\object{ROXs 39}         &   - &     - & 16 30 35.63  & -24 34 18.9	& 16 27 33.74    & -24 27 51.7     &  8.025   &     - &  5 & x  v  i  & B  & -\\
\object{Haro 1-14/c}     &   - &     - & 16 31 04.37  & -24 04 33.1	& 16 28 03.09    & -23 58 07.8     &  7.784   &     - &  9 & -- v  -- & B  & \object{HBC~644}\\
\object{Haro 1-14}       &   - &    69 & 16 31 05.17  & -24 04 40.1	& 16 28 03.89    & -23 58 14.9     &  8.576   &     - &  9 & -- v  -- & B  & \object{HBC~267}\\
\object{ROXs 42B}        &   - &     - & 16 31 15.02  & -24 32 43.7	& 16 28 13.13    & -24 26 19.1     &  8.671   &     - &  6 & x  v  -- & B  & -\\
\object{ROXs 42C}        &   - &     - & 16 31 15.75  & -24 34 02.2	& 16 28 13.83    & -24 27 37.7     &  7.129   &     - &  6 & x  v  -- & -  & -\\
\object{ROXs 43A/B}      &   - &     - & 16 31 20.12  & -24 30 05.2	& 16 28 18.29    & -24 23 41.0     &  6.729   &     - &  5 & x  v  i  & C  & \object{GWAYL 1}\\
\object{H$\alpha$ 71}    &   - &    71 & 16 31 30.88  & -24 24 40.0	& 16 28 29.15    & -24 18 16.4     &  7.900   &     - &  5 & -- v  i  & BC & \object{GWAYL 2}\\
\object{Haro 1-16}       &   - &    72 & 16 31 33.46  & -24 27 37.3	& 16 28 31.67    & -24 21 13.9     &  7.610   &     - &  5 & x  v  i  & O  & \object{GWAYL~3}, \object{ROXs~44}, \object{HBC~268}\\
\object{IRS 63}          &   - &     - & 16 31 35.66  & -24 01 29.5	& 16 28 34.42    & -23 55 06.3     &  9.219   &     - &  5 & -- -- i  & -  & \object{GWAYL 4}\\
\object{L1689-IRS 5}     & 204 &     - & 16 31 52.11  & -24 56 15.7	& 16 28 49.68    & -24 49 53.6     &  7.557   &     - &  5 & -- -- i  & -  & \object{GWAYL 5}\\
\object{H$\alpha$ 73}    & 206 &    73 & 16 31 54.42  & -25 03 49.3	& 16 28 51.82    & -24 57 27.4     &  9.899   &     - &  8 & -- v  i  & -  & -\\
\object{H$\alpha$ 74}    & 207 &    74 & 16 31 54.73  & -25 03 23.8	& 16 28 52.14    & -24 57 01.9     &  7.749   &     - &  8 & -- v  i  & -  & -\\
\object{ROXs 45D}        &   - &     - & 16 31 57.68  & -25 29 33.7	& 16 28 54.51    & -25 23 11.9     &  9.865   &     - &  6 & x  v  -- & -  & -\\
\object{ROXs 45E}        &   - &     - & 16 32 00.59  & -25 30 28.7	& 16 28 57.39    & -25 24 07.1     &  9.478   &     - &  6 & x  v  -- & -  & -\\
\object{ROXs 45F}        &   - &     - & 16 32 01.61  & -25 30 25.3	& 16 28 58.41    & -25 24 03.8     &  9.395   &     - &  6 & x  v  -- & -  & -\\
\object{H$\alpha$ 75}    &   - &    75 & 16 32 05.52  & -25 02 36.2	& 16 29 02.94    & -24 56 15.0     &  9.947   &     - & 10 & -- v  -- & U  & -\\
\object{DoAr 51}         &   - &    76 & 16 32 11.79  & -24 40 21.4	& 16 29 09.69    & -24 34 00.6     &  7.929   &     - &  6 & x  v  -- & -  & \object{ROXs~47A}, \object{HBC~647}\\
\object{L1689-IRS 7}     & 212 &     - & 16 32 21.05  & -24 30 35.8	& 16 29 19.15    & -24 24 15.7     &  8.620   &     - &  5 & -- -- i  & B  & \object{GWAYL 7}\\
\object{Haro 1-17}       &   - &    77 & 16 32 21.93  & -24 42 14.8	& 16 29 19.77    & -24 35 54.7     &  9.151   &     - &  9 & -- v  -- & -  & \object{HBC~648}\\
\object{Elias 45}        &   - &     - & 16 33 21.54  & -23 50 21.4	& 16 30 20.46    & -23 44 05.3     &  6.345   &     - & 11 & -- -- i  & -  & -\\
\noalign{\smallskip}        
\hline        
\noalign{\smallskip}        
\multicolumn{13}{l}{$Names\ adopted\ from\ \cite{bklt}\ are given without the leading `BKLT' and thus start with `J16'.$} \\
\noalign{\smallskip}        
\multicolumn{4}{l}{$A: observed with ADONIS/SHARP~II+$} & \multicolumn{3}{l}{$B: observed with BlackMAGIC$} & \multicolumn{5}{l}{$C: companion in 2MASS$}\\
\multicolumn{4}{l}{$E: 2MASS photometric quality flag E$} & \multicolumn{3}{l}{$L: $\mathrm{m}_\mathrm{K}>10.5$ (see text)$} & \multicolumn{5}{l}{$O: well observed$}\\
\noalign{\smallskip}        
\multicolumn{13}{l}{$Selection criteria: X-ray detection (x), visual spectra / H$\alpha$-emission (v), infrared excess (i) -- Selection \# refers to Table~\ref{selection}$}\\
\noalign{\smallskip}        
\hline
\end{array}
$$
\end{table*}


\section{Observations}

The principle part of the speckle observations (Table~\ref{Journal}) were carried out with the camera SHARP~I (System for High Angular Resolution Pictures) of the Max-Planck-Institute for Extraterrestrial Physics \citep{hofmann} mounted on the ESO New Technology Telescope (NTT) at La Silla, Chile. Further we obtained observations with BlackMAGIC \citep{herbst} on the 3.5\,m telescope at Calar Alto, Spain and with SHARP~II+ with the adaptive optics system ADONIS on the ESO 3.6\,m telescope at La Silla. In Table~\ref{Sample} objects observed with BlackMAGIC are marked with a `B' while those observed with ADONIS/SHARP~II+ are indicated by an `A'. All observations have been performed in the K-band at $2.2~\mu m$.

\begin{table}
\caption[]{Journal of observations}
\label{Journal}
$$
\begin{array}{p{0.33\linewidth}p{0.25\linewidth}p{0.32\linewidth}}
\hline
\noalign{\smallskip}
Camera     & Telescope & Date \\
\noalign{\smallskip}
\hline
\noalign{\smallskip}
SHARP~I                    & NTT, La Silla    & 2000, June 17 - 22\\
                           &                  & 2001, June 28 - July 4\\
BlackMAGIC                 & 3.5m, Calar Alto & 2000, June 22\\
SHARP~II+ / ADONIS         & 3.6m, La Silla   & 2000, June 5 - 6\\
\noalign{\smallskip}
$\Omega$-Cass (background) & 3.5m, Calar Alto & 2001, May 31 - June 1\\   
\noalign{\smallskip}
\hline
\end{array}
$$	
\end{table}

The cameras are equipped with an $256\times 256$ pixel NICMOS3 array.
To derive the exact pixel scale and orientation of the chips we took images of the \object{Galactic Center} and/or the \object{Orion Trapezium} during each observing campaign. We compared the instrumental positions of the stars with the very accurate coordinates given in \cite{genzel}, \cite{menten} and \cite{mccaughrean} by using the astrometric software {\it ASTROM}. In the case of the observations with BlackMAGIC no such calibration frames are available. Here we compared the position angle and separation of \object{H$\alpha$~71} given in \cite{koresko} with our result.

For each of the scientific targets we took between 500 and 1000 frames with an exposure time of 200 to 500~ms each to create the two fitscubes required for the data analysis (see~section~\ref{da}). We centered the primaries in one of the four quadrants of the detector and shifted the target after half of the frames had been taken to another quadrant. If no companion was visible below the primary we used the lower two quadrants. The advantage of this shifting is the exact measurement of the background, both at the same time in different areas of the chip, and in the same area at a different time. 

To analyse the data we also need speckle-images of stars that have no companions as PSF-references. In most cases we obtained time series of the star \object{HD~148352} with a spectral type of F2V \citep{houk} and a K-band magnitude of $6.511\pm~0.018\mathrm{\,mag}$ (2MASS). Another reference is the high proper-motion star \object{HD~148428} with a spectral type of K0III+G \citep{houk} and a K-band magnitude of $5.925\pm~0.024\mathrm{\,mag}$ (2MASS). These two stars are located in the foregound of the southern wing of \object{L1688} and they are comparable to our brightest targets in the K-band.  An additional reference used during the observations with BlackMAGIC is the G1 main-sequence star \object{HD~147681}. Its brightness is $7.508\pm~0.023\mathrm{\,mag}$ in the K-band (2MASS). The references for the objects observed with SHARP~II+ are the single targets themselves.


\section{Data Analysis}
\label{da}

\begin{figure*}
\centering
\includegraphics[width=18.1cm]{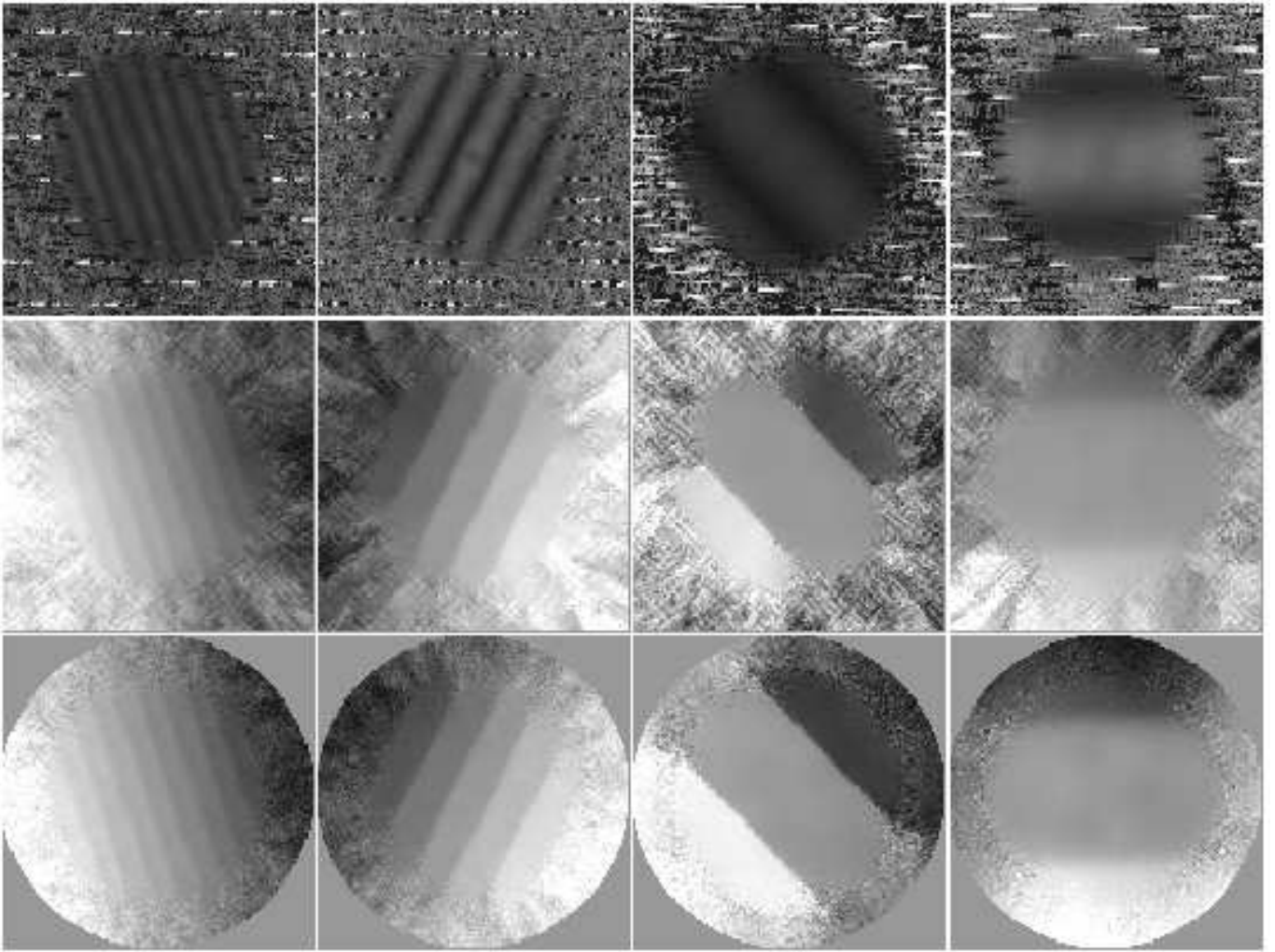}
\caption{The visibility, the Knox-Thompson phase, and the bispectrum phase (from top to bottom) of the sources IRS~3, ROXs~31, VSSG~5, and VSSG~11 (from left to right) as reconstructed by our software. The increasing gap between the maxima in the visibility and the decreasing number of steps in the phases clearly indicate a decrease in separation: 0.663'', 0.396'', 0.148'', and 0.107''. VSSG~11 is an example of an object falling below the diffraction limit of the telescope, i.e.~we are not able to decide whether it is a binary or an elongated structure. The spatial vector between the two components of a binary is perpendicular to the stripes in the visibility and the phases. The overall gradient of the phases eliminates the $180^{\circ}$ ambiguity. We get position angles of $115.5^{\circ}$, $251.3^{\circ}$, $133.9^{\circ}$, and $180.1^{\circ}$. The flux ratio (0.323, 0.655, 0.873, and 0.584) can be determined by the amplitude of the sinusoidal wave in the visibility and the transition between the steps in the phases. The smaller the flux ratio the smaller the height of the steps. The  equally spaced horizontal stripes in the visibilities are artefacts, probably from an interference with the readout electronics.}
\label{examples}
\end{figure*}

\begin{figure*}
\includegraphics[width=12cm]{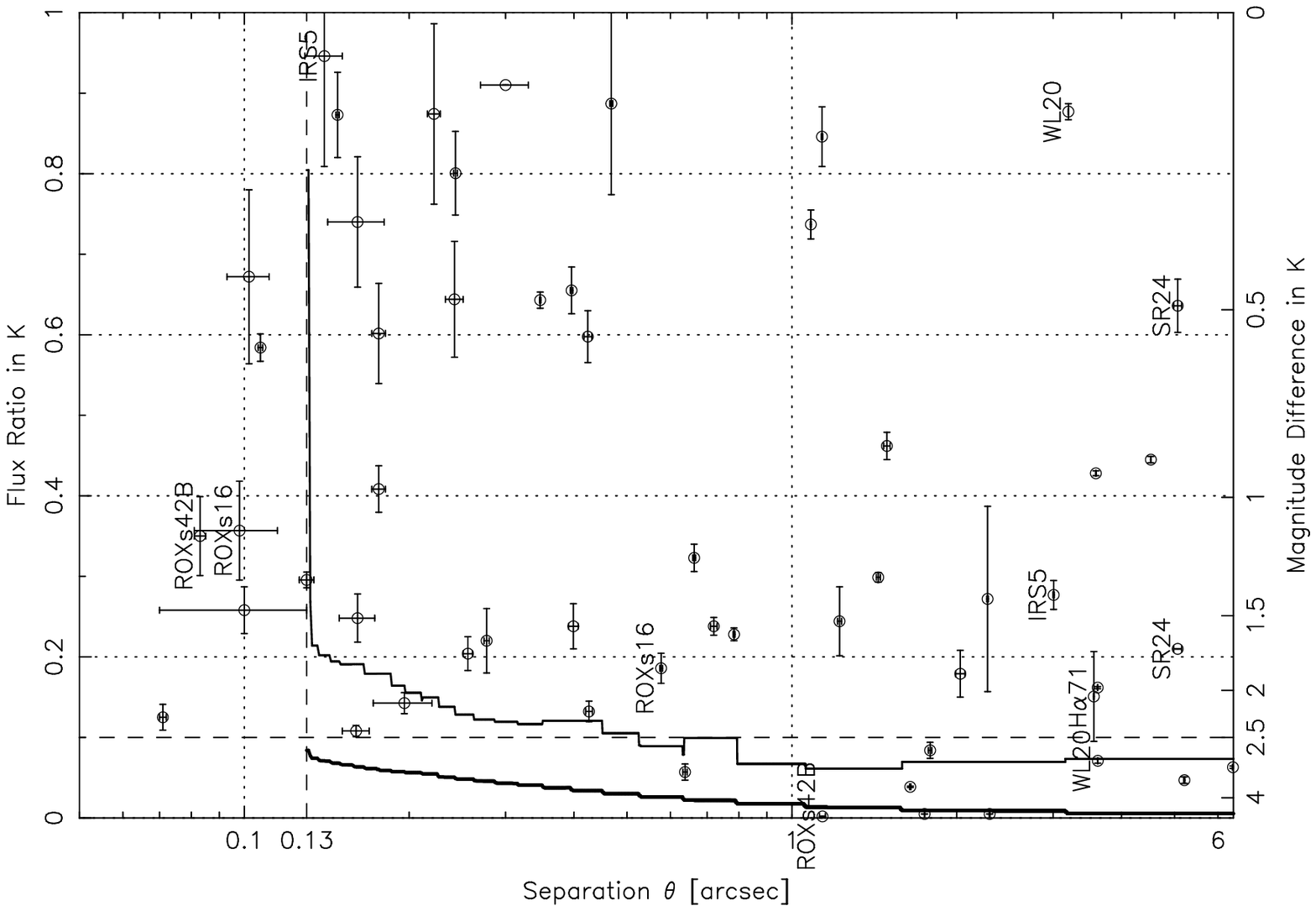}
\caption{Results of our multiplicity survey in a plot of flux ratio or magnitude difference vs. binary star separation. The data points mark the detected companion stars. If a companion is a component of a triple star it is labeled with the name of the system. The thick line is the average, and the thin line the worst sensitivity for undetected companions. The dashed vertical line at 0.13'' shows the diffraction limit for a 3.5~m telescope at K. This is the limit for unambiguous identification of binary stars. The dashed horizontal line shows the completeness limit in flux ratio for the whole survey.}
\label{results}
\end{figure*}

We used our program {\it speckle} which has already been used for the surveys in Taurus-Auriga \citep{koehler1}, Scorpius-Centaurus \citep{koehler2}, and Chamaeleon \citep{koehler3}. In this program, the modulus of the complex visibility (i.e., the Fourier transform of the object brightness distribution) is determined from power spectrum analysis, and the phase is computed using the Knox-Thompson algorithm \citep{knox} and from the bispectrum \citep{lohmann}. For a more detailed description see \cite{koehler2}. A few examples are presented in Fig.~\ref{examples}.

If the object turns out to be a binary or multiple star, we obtain the position angle, separation and brightness ratio of the components from a multidimensional least-square fit. Our program tries to minimize the difference between modulus and phase computed from a model binary and the observational data by varying the binary parameters. Fits to different subsets of the data give an estimate for the standard deviation of the binary parameters.

For the search of wide and faint companions we coadded the frames within one fitscube. This increases the sensitivity, because signals are amplified while the noise is reduced. On the other hand this merging process would introduce the atmospheric distortion again that has been `frozen' by the short integration time of speckle images. The technique of shift-and-add partially circumvents this problem. Here the frames are centered before they are summed up. This centering can be done either on the brightest pixel or the center of light. Afterwards, these pictures can be analysed with standard tools, like {\it daophot}. 

If the object appears unresolved, we compute the maximum brightness ratio of a companion that could be hidden in the noise of the data. The principle is to determine how far the data deviate from the nominal result for a point source (modulus=1, phase=0) and to interpret this deviation as caused by a companion. The procedure is repeated for different position angles and the maximum is used as an upper limit for the brightness ratio of an undetected companion \citep{leinert}. In Table~\ref{Singles} we list the values at a distance of 0.15~arcsec and 0.50~arsec from the primary. After subtraction of the companion(s) the first value is also calculated for double or multiple stars as an indicator for the quality of the fit (Table~\ref{Doubles}).


\section{Results}

\subsection{Uncorrected Data}

In Table~\ref{Singles} and Table~\ref{Doubles} we list our results. Objects also observed in other near-infrared high-resolution studies are identified. In total, among the 158 targets of our sample, we find up to separations of 6.4'' 45 binaries, 5 triple systems (\object{ROXs~16}, \object{WL~20}, \object{ROXs~42B}, \object{L1689-IRS~5}, and \object{SR~24}), and no quadruples. The flux ratio or magnitude difference vs. the separation of these systems is plotted in Fig.~\ref{results}.

\begin{table*}
\caption[]{Upper limits for the relative brightness of an undetected companion to the unresolved stars in our survey, measured at 0.15'' and 0.50''.}
\label{Singles}$$
\begin{array}
           {p{0.12\linewidth}p{0.11\linewidth}p{0.04\linewidth}p{0.04\linewidth}p{0.13\linewidth}
            p{0.14\linewidth}p{0.14\linewidth}p{0.04\linewidth}p{0.04\linewidth}p{0.11\linewidth}}
\hline
\noalign{\smallskip}
Object & Date & 0.15'' & 0.50'' & References$^\mathrm{*}$ & Object & Date & 0.15'' & 0.50'' & References$^\mathrm{*}$\\
\noalign{\smallskip}
\hline
\noalign{\smallskip}
H$\alpha$ 16	& 2001, July 4   & 0.09  & 0.03 & A1			       & IRS 32          & 2000, June 22 & 0.08  & 0.04 &\\
H$\alpha$ 22	& 2001, July 4   & 0.08  & 0.03 &			       & VSSG 24         & 2001, June 29 & 0.07  & 0.03 &\\
SR 22	 	& 2000, June 17  & 0.07  & 0.07 & B2			       & IRS 32b         & '91, Aug./'92, July   & \multicolumn{2}{l}{0.5\ \ (0.02'')} & S2\\
SR 1	 	& 2000, June 6   & 0.09  & 0.03 &			       & ROXs 20A        & 2000, June 20 & 0.10  & 0.10 &\\
SR 8	 	& 2001, July 4   & 0.14  & 0.07 &			       & ROXs 20B        & 2000, June 20 & 0.09  & 0.07 &\\
Elias 12	& 2000, June 17  & 0.03  & 0.02 &			       & H$\alpha$ 47    & 2000, June 21 & 0.09  & 0.04 &\\
H$\alpha$ 24	& 2001, July 4   & 0.12  & 0.05 &			       & WL 5            & 2000, June 20 & 0.18  & 0.13 &\\
IRS 8	 	& 2000, June 21  & 0.05  & 0.02 &			       & IRS 42          & 2000, June 21 & 0.06  & 0.03 & C, S2\\
IRS 9	 	& 2000, June 20  & 0.09  & 0.06 & C			       & WL 6            & 2001, July 1  & 0.11  & 0.05 &\\
ROXs 3	 	& 2000, June 17  & 0.06  & 0.04 & S2			       & VSSG 22         & 2000, June 20 & 0.05  & 0.04 & C\\
VSS 23	 	& 2000, June 17  & 0.05  & 0.05 & B2			       & H$\alpha$ 49    & 2000, June 20 & 0.06  & 0.06 &\\
IRS 11	 	& 2001, July 3   & 0.09  & 0.04 &			       & GY 262          & 2000, June 21 & 0.10  & 0.04 &\\
SR 4	 	& 1990, Aug. 7   & 0.05  & 0.04 & C, G2, S2		       & IRS 43          & 2001, June 29 & 0.07  & 0.07 & C, S2\\
GSS 20	 	& 2000, June 17  & 0.05  & 0.04 & A1, C		       	       & VSSG 18         & 2000, June 20 & 0.12  & 0.06 &\\
Chini 8	 	& 2001, June 30  & 0.16  & 0.04 &			       & GY 284          & 2001, July 1  & 0.08  & 0.04 &\\
DoAr 21         & 1990, July 9   & 0.06  & 0.06 & A1, C, G2, S2	               & J162730-244726  & 2001, June 29 & 0.07  & 0.05 &\\
VSSG 19	 	& 2000, June 21  & 0.09  & 0.04 &			       & GY 292          & 2000, June 20 & 0.03  & 0.03 & B2\\
Chini 11	& 2001, July 3   & 0.13  & 0.06 &			       & H$\alpha$ 50    & 2000, June 21 & 0.05  & 0.02 &\\
SR 3	 	& 2000, June 6   & 0.05  & 0.04 & C, S2  		       & IRS 48          & 2000, June 20 & 0.07  & 0.02 & C, S2\\
GSS26	 	& 2000, June 20  & 0.06  & 0.03 & C			       & IRS 50          & 2000, June 20 & 0.10  & 0.04 & C\\
SKS 1-7	 	& 2001, June 30  & 0.05  & 0.02 &			       & IRS 49          & 2000, June 21 & 0.04  & 0.02 & C, S2\\
GSS29	 	& 2000, June 17  & 0.04  & 0.04 & C, S2  		       & ROXs 30B        & 2000, June 21 & 0.07  & 0.03 & A1, B2\\
DoAr 24         & 1990, Aug. 7   & 0.09  & 0.07 & B2, C, G2 		       & ROXs 30C        & 2000, June 21 & 0.08  & 0.02 & A1\\
VSSG1	 	& 2000, June 20  & 0.04  & 0.03 & C			       & H$\alpha$ 52    & 2000, June 21 & 0.04  & 0.02 & B2, S2\\
J162621-241544  & 2001, June 29  & 0.08  & 0.03 &			       & IRS 56          & 2001, July 3  & 0.11  & 0.06 & S2\\
Elias 21	& 2000, June 20  & 0.04  & 0.02 & C			       & SR 10           & 2000, June 21 & 0.04  & 0.02 & C, R3, S2\\
GSS 30 - IRS 2  & 2000, June 20  & 0.10  & 0.06 &			       & H$\alpha$ 58    & 2001, June 29 & 0.12  & 0.03 &\\
LFAM 3	 	& 2001, June 30  & 0.10  & 0.04 & C			       & J162800-245340  & 2001, June 30 & 0.05  & 0.03 &\\
DoAr 25	 	& 2000, June 6   & 0.05  & 0.03 & C			       & VSS 35          & 2001, July 4  & 0.10  & 0.05 & R3$^\mathrm{**}$\\
GSS 32	 	& 2000, June 18  & 0.03  & 0.03 & R3, S2		       & J162813-243249  & 2001, July 1  & 0.09  & 0.04 &\\
Elias 24	& 2000, June 17  & 0.10  & 0.03 & C			       & H$\alpha$ 60    & 2000, June 21 & 0.03  & 0.02 & S2\\
H$\alpha$ 33	& 2001, July 4   & 0.06  & 0.03 &			       & ISO-Oph 195     & 2001, June 29 & 0.05  & 0.04 &\\
GY 33	 	& 2001, June 30  & 0.06  & 0.02 &			       & SR 20 W (GWAYL) & 2001, July 3  & 0.08  & 0.04 &\\
S1	 	& 2000, June 21  & 0.05  & 0.02 & A1, C, S2$^\mathrm{**}$, R3$^\mathrm{**}$ & VSS 38 & 2000, June 17 & 0.03  & 0.02&\\
J162636-241554  & 2001, July 3   & 0.12  & 0.05 &   	     		       & H$\alpha$ 63    & 2000, June 17 & 0.07  & 0.05 & S2\\
WL 8	 	& 2001, June 29  & 0.09  & 0.04 &			       & VSS 42          & 2001, July 4  & 0.07  & 0.02 & R3\\
GY 112	 	& 2001, June 30  & 0.19  & 0.05 &			       & IRAS 64a        & 2000, June 21 & 0.06  & 0.02 &\\
GSS39	 	& 2000, June 20  & 0.06  & 0.03 & C			       & VSS 41          & 2001, July 4  & 0.04  & 0.03 &\\
Haro 1-8	& 2000, June 22  & 0.04  & 0.02 &   	     		       & Elias 41        & 2001, July 3  & 0.19  & 0.06 &\\
H$\alpha$ 40	& 2001, July 3   & 0.06  & 0.03 &   	     		       & H$\alpha$ 67    & 2001, July 4  & 0.14  & 0.04 & S2\\
VSSG 10	 	& 2001, July 3   & 0.10  & 0.05 &   	     		       & ROXs 39         & 2000, June 22 & 0.09  & 0.06 & A1\\
VSSG 7	 	& 2001, June 30  & 0.14  & 0.03 &   	     		       & Haro 1-14/c     & 2000, June 22 & 0.06  & 0.06 & B2\\
J162656-241353 	& 2001, June 30  & 0.04  & 0.02 &           		       & Haro 1-14       & 2000, June 22 & 0.05  & 0.03 & B2,G2\\
VSSG 8	 	& 2001, June 29  & 0.06  & 0.03 &   	     		       &	         & 2001, July 4  & 0.07  & 0.04 &\\
H$\alpha$ 44	& 2001, July 4   & 0.10  & 0.06 &   	     		       & Haro 1-16       & 1990, Aug. 6  & 0.05  & 0.05 & B2, G2, R3, S2\\
WL16	 	& 2000, June 18  & 0.06  & 0.03 & C, S2  		       & IRS 63          & 2001, July 4  & 0.08  & 0.03 &\\
VSSG 9	 	& 2001, July 3   & 0.06  & 0.03 &			       & H$\alpha$ 73    & 2001, July 2  & 0.09  & 0.07 & S2\\
GY 193	 	& 2001, June 30  & 0.09  & 0.03 &			       & H$\alpha$ 74    & 2001, July 2  & 0.07  & 0.04 & B2\\
GY 194	 	& 2001, June 30  & 0.10  & 0.04 &			       & ROXs 45D        & 2001, July 2  & 0.07  & 0.02 &\\
VSSG 21	 	& 2001, July 3   & 0.12  & 0.05 &			       & ROXs 45E        & 2001, July 2  & 0.10  & 0.04 &\\
J162708-241204  & 2001, June 30  & 0.05  & 0.02 &   	     		       & ROXs 45F        & 2001, July 2  & 0.08  & 0.04 &\\
WL 10	 	& 2000, June 21  & 0.08  & 0.03 &   	     		       & H$\alpha$ 75    & 2001, July 1  & 0.06  & 0.02 &\\
Elias 29	& 2000, June 21  & 0.03  & 0.02 & C, S2         	       & L1689 - IRS 7   & 2000, June 22 & 0.05  & 0.04 &\\
 	 	& 2000, June 21  & 0.09  & 0.04 &   	     		       & Haro 1-17       & 2001, July 2  & 0.05  & 0.02 &\\
GY 224	 	& 2000, June 22  & 0.09  & 0.05 &   	     		       & Elias 45        & 2001, July 3  & 0.13  & 0.05 &\\
\noalign{\smallskip}
\hline
\noalign{\smallskip}
\multicolumn{10}{l}{$Names\ adopted\ from\ \cite{bklt}\ are given without the leading `BKLT' and thus start with `J16'.$} \\
\noalign{\smallskip}        
\multicolumn{5}{l}{^\mathrm{*}$: references are given in Table~\ref{Barsony} in the appendix$} &
\multicolumn{5}{l}{^\mathrm{**}$: additional lunar occultation companions $} \\ 
\noalign{\smallskip}
\hline
\end{array}
$$
\end{table*}

\begin{table*}
\caption[]{The double and multiple stars in our sample. Given are the position angles, the separations, and the flux ratios. The upper limit for the relative brightness of an additional undetected companion at the diffraction limit is provided in the seventh column.}
\label{Doubles}$$
\begin{array}
          {p{0.12\linewidth}p{0.05\linewidth}p{0.14\linewidth}p{0.09\linewidth}p{0.11\linewidth}p{0.12\linewidth}
          p{0.06\linewidth}p{0.23\linewidth}}
\hline
\noalign{\smallskip}
Object &  & Date & PA [deg] & Separation [''] & Flux Ratio & 0.15'' & Remarks$^\mathrm{*}$\\
\noalign{\smallskip}
\hline
\noalign{\smallskip}
H$\alpha$ 18 	&       & 2001, July 4	     & 82.3 $\pm$ 0.1   & 1.083 $\pm$ 0.002 & 0.737 $\pm$ 0.018 & 0.15 & \\
H$\alpha$ 19 	&       & 2001, July 4	     & 262.9 $\pm$ 0.1  & 1.491 $\pm$ 0.020 & 0.462 $\pm$ 0.017 & 0.05 & \\
Haro 1-4 	&       & 1990, July 9	     & 27 $\pm$ 1       & 0.72  $\pm$ 0.01  & 0.238 $\pm$ 0.011 & 0.05 & G2\\
H$\alpha$ 21 	&       & 2001, July 4	     & 57.6 $\pm$ 1.6   & 0.161 $\pm$ 0.019 & 0.740 $\pm$ 0.081 & 0.07 & \\
SR 2 	 	&       & 2000, June 5	     & 122.4 $\pm$ 0.6  & 0.222 $\pm$ 0.006 & 0.874 $\pm$ 0.112 & 0.06 & G2\\
ROXs 2 	 	&       & 2000, June 22	     & 345.5 $\pm$ 1.4  & 0.424 $\pm$ 0.007 & 0.598 $\pm$ 0.032 & 0.05 & B2, C\\
IRS 2 	 	&       & 2000, June 17	     & 78.6 $\pm$ 0.4   & 0.426 $\pm$ 0.006 & 0.132 $\pm$ 0.013 & 0.10 & B2, C\\
J162538-242238 	&       & 2001, July 4	     & 170.2 $\pm$ 0.5  & 1.788 $\pm$ 0.013 & 0.084 $\pm$ 0.010 & 0.06 & \\
IRS 3 	 	&       & 2001, June 29	     & 115.5 $\pm$ 0.6  & 0.663 $\pm$ 0.004 & 0.323 $\pm$ 0.017 & 0.04 & \\
ROXs 5 	 	&       & 2000, June 22	     & 327.3 $\pm$ 1.7  & 0.176 $\pm$ 0.005 & 0.408 $\pm$ 0.029 & 0.03 & A1\\
ROXR1-12 	&       & 2001, June 30	     & 18.5 $\pm$ 2.9   & 0.102 $\pm$ 0.009 & 0.672 $\pm$ 0.108 & 0.12 & \\
H$\alpha$ 26 	&       & 2001, July 4	     & 25.8 $\pm$ 0.5   & 1.135 $\pm$ 0.004 & 0.846 $\pm$ 0.037 & 0.11 & \\
DoAr 22 	&       & 2001, July 2	     & 258.9 $\pm$ 0.2  & 2.297 $\pm$ 0.004 & 0.005 $\pm$ 0.000 & 0.03 & \\
H$\alpha$ 28 	&       & 2001, June 29	     & 357.8 $\pm$ 0.1  & 5.209 $\pm$ 0.013 & 0.047 $\pm$ 0.004 & 0.09 & \\
DoAr 24E 	&       & 1990, July 9	     & 150 $\pm$ 1      & 2.03  $\pm$ 0.04  & 0.179 $\pm$ 0.029 & 0.05 & A1, C, G2, S2\\
ROXs 12 	&       & 2001, July 2	     & 10.3 $\pm$ 0.1   & 1.747 $\pm$ 0.002 & 0.005 $\pm$ 0.000 & 0.06 & \\
VSSG 27 	&       & 2000, June 20	     & 66.8 $\pm$ 0.5   & 1.222 $\pm$ 0.010 & 0.244 $\pm$ 0.043 & 0.05 & C\\
H$\alpha$ 35 	&       & 2001, July 4	     & 132.2 $\pm$ 0.1  & 2.277 $\pm$ 0.007 & 0.272 $\pm$ 0.115 & 0.10 & \\
H$\alpha$ 37 	&       & 2000, June 20	     & 65 $\pm$ 2       & 0.16  $\pm$ 0.01  & 0.108 $\pm$ 0.007 & 0.10 & not seen by C, PA mod 180$^{\circ}$\\
GSS 37 	 	&       & 2000, June 18	     & 69.5 $\pm$ 0.3   & 1.438 $\pm$ 0.012 & 0.299 $\pm$ 0.006 & 0.05 & C\\
VSSG 11 	&       & 2001, July 1	     & 180.1 $\pm$ 0.6  & 0.107 $\pm$ 0.001 & 0.584 $\pm$ 0.017 & 0.04 & \\
ROXs 16 	& Aa-Ab & 2000, June 21	     & 24.2 $\pm$ 7.5   & 0.098 $\pm$ 0.017 & 0.357 $\pm$ 0.061 & 0.05 & \\
 	 	& Aa-B  &	             & 105.4 $\pm$ 0.6  & 0.577 $\pm$ 0.003 & 0.186 $\pm$ 0.019 &      & A1, C\\
WL18 	 	&       & 2000, June 22	     & 292.4 $\pm$ 0.2  & 3.617 $\pm$ 0.001 & 0.162 $\pm$ 0.001 & 0.04 & \\
VSSG 3 	 	&       & 2000, June 21	     & 53.8 $\pm$ 0.5   & 0.243 $\pm$ 0.002 & 0.801 $\pm$ 0.052 & 0.07 & C\\
VSSG 5 	 	&       & 2001, June 30	     & 133.9 $\pm$ 1.3  & 0.148 $\pm$ 0.001 & 0.873 $\pm$ 0.053 & 0.04 & \\
GY 156 	 	&       & 2000, June 21	     & 201.9 $\pm$ 1.8  & 0.161 $\pm$ 0.012 & 0.248 $\pm$ 0.030 & 0.07 & \\
SR 24 	 	& S-N   & 1999, Apr. 17	     & 349.4 $\pm$ 1.3  & 5.065 $\pm$ 0.086 & 0.636 $\pm$ 0.033 & 0.06 & G2, S2 (flux limit at 0.02'')\\
 	 	& Na-Nb & 1991, Aug. 19	     & 84	        & 0.197 $\pm$ 0.020 & 0.21	        &      & C, S2\\
Elias 30 	&       & 2000, June 21	     & 175.6 $\pm$ 0.2  & 6.388 $\pm$ 0.013 & 0.063 $\pm$ 0.002 & 0.06 & S2, R3, not seen by C\\
WL 20 	 	& A-B   & 2001, July 1	     & 269.9 $\pm$ 0.1  & 3.198 $\pm$ 0.000 & 0.877 $\pm$ 0.010 & 0.06 & \\
 	 	& A-C   &		     & 232.3 $\pm$ 0.1  & 3.619 $\pm$ 0.001 & 0.071 $\pm$ 0.003 &      & \\
WL 4 	 	&       & 2000, June 20	     & 284.2 $\pm$ 2.3  & 0.176 $\pm$ 0.005 & 0.602 $\pm$ 0.062 & 0.06 & not seen by C \\
SR 12 	 	&       & '86, Jan. / '91, Aug. & 85	        & 0.300 $\pm$ 0.030 & 0.91              & 0.33 & C, S2 (flux limit at 0.02'')\\
VSSG 25 	&       & 2000, June 20	     & 173.3 $\pm$ 0.3  & 0.468 $\pm$ 0.003 & 0.887 $\pm$ 0.113 & 0.10 & C \\
IRS 44 	 	&       & 2001, June 30	     & 246.6 $\pm$ 5.1  & 0.256 $\pm$ 0.005 & 0.204 $\pm$ 0.021 & 0.2  & not seen by C and S2, bad s/n\\
VSSG 17 	&       & 2000, June 21	     & 260.2 $\pm$ 0.8  & 0.242 $\pm$ 0.009 & 0.644 $\pm$ 0.072 & 0.04 & C \\
IRS 51 	 	&       & 2000, June 20	     & 9.6 $\pm$ 0.3    & 1.645 $\pm$ 0.005 & 0.039 $\pm$ 0.001 & 0.07 & not seen by C and S2\\
SR 9 	 	&       & 2001, July 3	     & 353.3 $\pm$ 0.5  & 0.638 $\pm$ 0.006 & 0.057 $\pm$ 0.010 & 0.17 & B2, G2\\
GY 371 	 	&       & 2001, June 30	     & 198.1 $\pm$ 0.3  & 0.347 $\pm$ 0.001 & 0.643 $\pm$ 0.010 & 0.06 & \\
VSSG 14 	&       & 2000, June 18	     & 83.6 $\pm$ 1.5   & 0.130 $\pm$ 0.004 & 0.296 $\pm$ 0.010 & 0.04 & S2, R3, not seen by C\\
ROXs 31 	&       & 2001, June 29	     & 251.3 $\pm$ 0.2  & 0.396 $\pm$ 0.002 & 0.655 $\pm$ 0.029 & 0.05 & A1, C, S2\\
GY 410 	 	&       & 2000, June 20	     & 277.0 $\pm$ 1.4  & 0.196 $\pm$ 0.024 & 0.143 $\pm$ 0.013 & 0.05 & \\
H$\alpha$ 59 	&       & 2001, July 4	     & 103.2 $\pm$ 2.4  & 0.100 $\pm$ 0.030 & 0.258 $\pm$ 0.029 & 0.07 & \\
J162812-245043  &       & 2001, July 3	     & 101.7 $\pm$ 0.1  & 3.591 $\pm$ 0.001 & 0.428 $\pm$ 0.003 & 0.15 & \\
SR 20 	 	&       & 1990, July 9	     & 225 $\pm$ 5      & 0.071 $\pm$ 0.001 & 0.125 $\pm$ 0.016 & 0.03 & G2, R3, S2, not seen by C\\
V 853 Oph 	&       & 1990, Aug. 7	     & 96 $\pm$ 2       & 0.399 $\pm$ 0.008 & 0.238 $\pm$ 0.028 & 0.14 & C, G2, S2 (triple$^\mathrm{**}$)\\
ROXs 42B 	& Aa-Ab & 2001, July 1 	& 157.9 $\pm$ 1.7  & 0.083 $\pm$ 0.002 & 0.350 $\pm$ 0.049 & 0.06 & R3, S2, not seen by A1\\
 	 	& A-B   &		& 268.0 $\pm$ 0.3  & 1.137 $\pm$ 0.014 & 0.002 $\pm$ 0.001 &	  & \\
ROXs 42C 	&       & 2001, July 1	     & 151.0 $\pm$ 0.7  & 0.277 $\pm$ 0.003 & 0.220 $\pm$ 0.040 & 0.05 & B2, G2\\
ROXs 43A/B 	&       & 2001, July 4	     & 11.9 $\pm$ 0.1   & 4.523 $\pm$ 0.004 & 0.445 $\pm$ 0.004 & 0.07 & A1, G2, S2 (quadruple$^\mathrm{**}$)\\
H$\alpha$ 71 	&       & 2000, June 22	     & 35.0 $\pm$ 1.4   & 3.560 $\pm$ 0.006 & 0.151 $\pm$ 0.056 & 0.04 & S2\\
L1689 - IRS 5   & A-Ba  & 2001, July 2	&  241.2 $\pm$ 0.1 & 3.006 $\pm$ 0.009 & 0.277 $\pm$ 0.018 & 0.05 & \\
 	        & Ba-Bb &			& 84.4 $\pm$ 6.1   & 0.140 $\pm$ 0.011 & 0.946 $\pm$ 0.137 &	  & \\
DoAr 51 	&       & 2001, July 2	     & 79.3 $\pm$ 0.2   & 0.784 $\pm$ 0.003 & 0.228 $\pm$ 0.008 & 0.06 & B2 \\
\noalign{\smallskip}
\hline
\noalign{\smallskip}
\multicolumn{8}{l}{$Names\ adopted\ from\ \cite{bklt}\ are given without the leading `BKLT' and thus start with `J16'.$} \\
\noalign{\smallskip}
\multicolumn{4}{l}{^\mathrm{*}$: references are given in Table~\ref{Barsony} in the appendix$} &
\multicolumn{4}{l}{^\mathrm{**}$: additional spectroscopic or lunar occultation companions $} \\ 
\noalign{\smallskip}
\hline
\end{array}
$$
\end{table*}

\subsection{Completeness}

The sensitivity of our survey as a function of the separation (see Fig.~\ref{results}) depends on factors such as atmospheric conditions at the time of the observations or the brightness of the target star. Since we derive for each dataset with our reduction method the maximum brightness ratio of a possible undetected companion (see Table~\ref{da}), we can continuously monitor the quality of our data. At the diffraction limit we reached in 85\% of the observations our quality criterion of a flux ratio $\leq 0.1$ ($\geq 2.5~\mathrm{mag}$) in the K-band. Twenty-two observations are not quite sensitive enough to fit this request. The maximum brightness of an undetected companion at the diffraction limit varies here between 0.11 and 0.19. In the case of \object{IRS~44} where the data are very noisy, we provide in Table~\ref{Doubles} the flux ratio of the detected companion ($\sim 0.2$) at a separation of 0.26'' as upper limit for the brightness of an undeteced companion.

Based on the surface density of companions found in Fig.~\ref{results} at separations larger than 0.13'' in the range between the requested flux ratio of 0.1 and the detection limits of the twenty-two measurements described above, the probability to have missed one companion is 40\%. Since the real sensitivity deficit is only relevant for separations below 1'', this estimate represents an upper limit. We are thus confident, that we have found all companions with a magnitude difference $\leq 2.5~\mathrm{\,mag}$.

\subsection{Lower Separation Limit}
\label{sr}

The lower limit of 0.13'' is given by the diffraction limit $\lambda/D$ of a 3.5\,m telescope in K. Nevertheless, it is possible to detect under good circumstances companions down to a separation of ${1\over 2}\lambda/D$, where the first minimum of the modulus of the complex visibility can be seen. In these cases it is not longer possible to definitely distinguish between an elongated structure and a binary star. Fig.~\ref{results} shows that we actually found such candidates: \object{ROXR1-12}, \object{VSSG~11}, \object{ROXs~16}, \object{H$\alpha$~59} and \object{ROXs~42B}. Also the close companion of \object{SR~20} would fall below our diffraction limit. It was detected below the diffraction limit of the Hale 5~m Telecope of Palomar Observatory by \cite{ghez}.

\subsection{Background}
\label{back}

\begin{figure}
\centering
\includegraphics[height=8.8cm,angle=270]{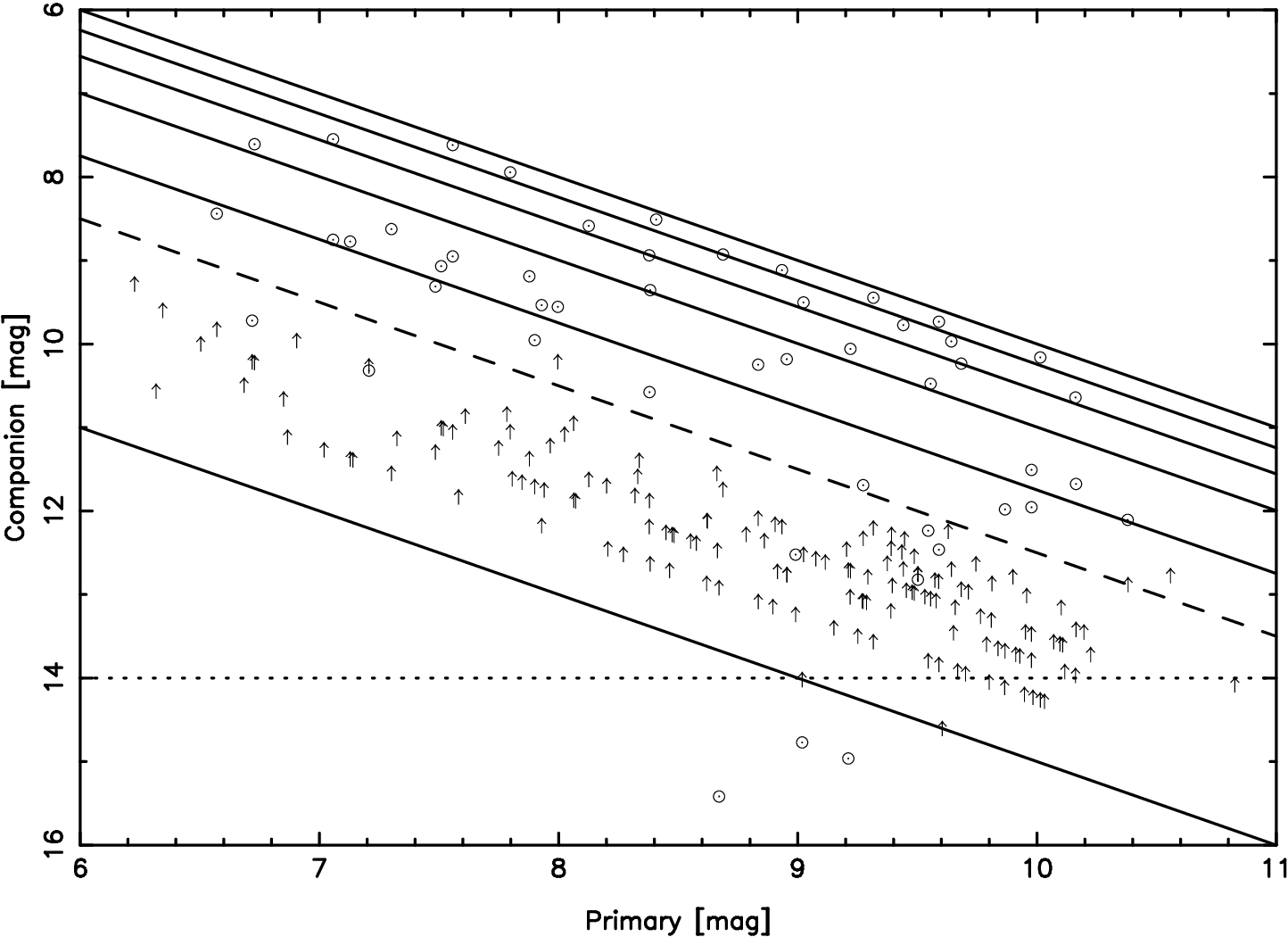}
\caption{The brightness of detected ($\odot$) and the upper limit for non-detected companions at a separation of 0.5'' ($\uparrow$) vs. the brightness of the primaries. The diagonal lines indicate flux ratios of 0.01, 0.1 (=completeness, dashed), 0.2, 0.4, 0.6, 0.8, and 1.0. The dotted horizontal line gives the magnitude used for the background determination.}
\label{res2}
\includegraphics[height=8.8cm,angle=270]{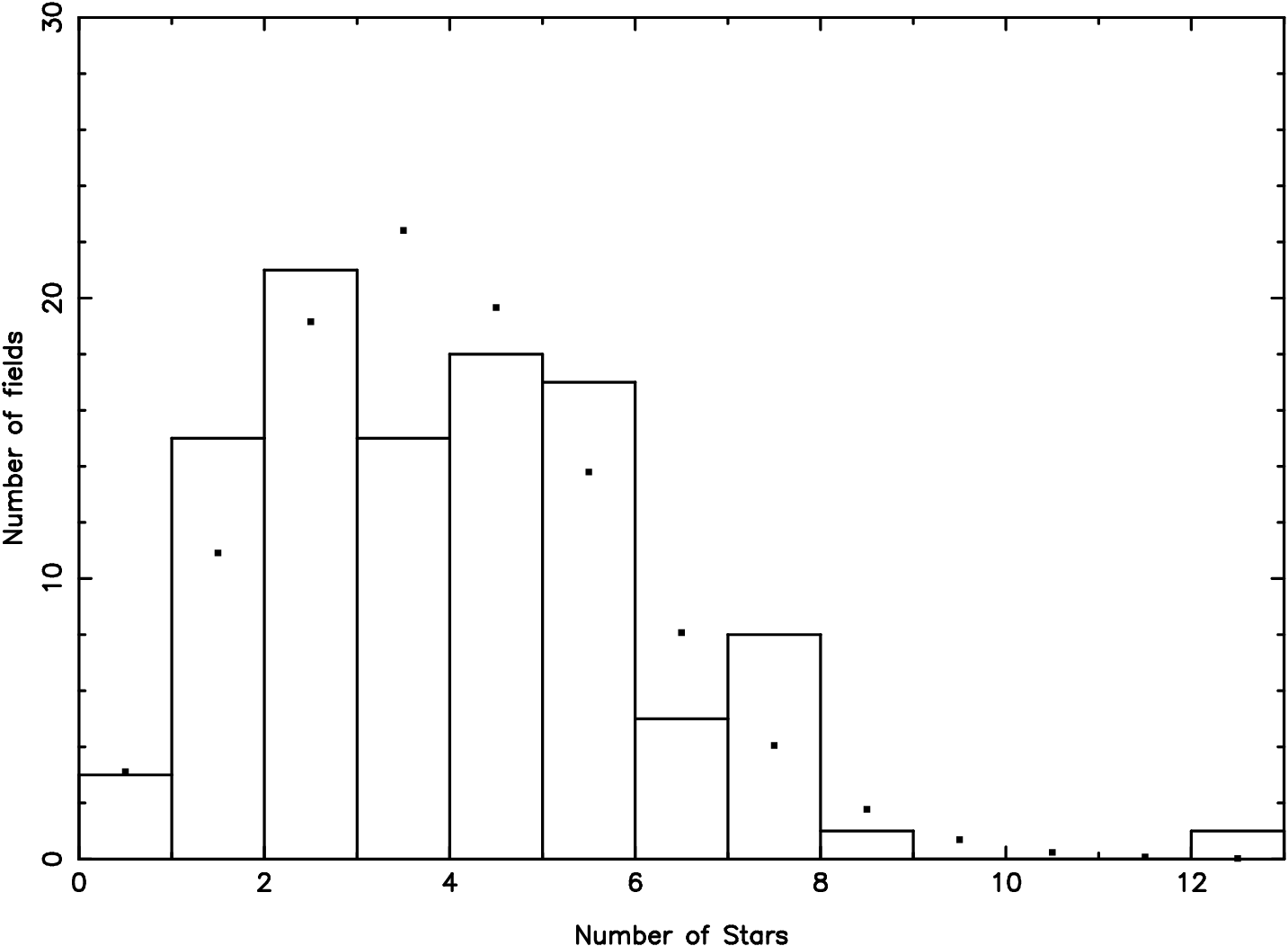}
\caption{Background statistics: In the 104 fields all stars down to a brightness limit of 14~mag are included. The dots represent a Poisson distribution with the mean value of $\approx 3.5$.}
\label{bs14}
\end{figure}

Since it is not possible with our data to distinguish between gravitationally bound companions and mere chance projections of background stars, it is necessary to quantify this bias. Therefore, we analysed 104 fields covering $\approx 6.5\mathrm{\,arcmin}^{-2}$ each, that are centered around 24 infrared sources of our sample, \object{ISO-Oph~13} and \object{VSS~28} (see Fig.~\ref{cloudcomplex}). These fields were created by mosaicing images obtained with the infrared camera $\Omega$-Cass in the K- or $\mathrm{K}_s$-band (see Table~\ref{Journal}). $\Omega$-Cass was mounted on the 3.5\,m telescope at Calar Alto, Spain. After excluding the central region with a radius of 6.4~arcsec corresponding to the largest separation found in our sample, we divided each mosaic into four equal fields.

Although we detected three companions with a K-band magnitude around 15 (see Fig.~\ref{res2}), these were found in the shift-and-add images and are thus not representative for the detection limit of our survey. The upper limits provided by the speckle software for non-detected companions are much better suited for this purpose. As shown in Fig.~\ref{res2} they correspond to $\mathrm{m}_\mathrm{K}=14~\mathrm{mag}$ for the fainter primaries.

The results of counting the stars down to the 14th magnitude in each field is plotted in Fig.~\ref{bs14}. The histogram can be fitted by a Poisson distribution with a mean value of $\approx 3.5$, which corresponds to an absolute value of $1.5\cdot10^{-4}\mathrm{\,arcsec}^{-2}$. Defining the area within $16^h25^m\dots 16^h30^m$ in right ascension and $-25^{\circ}\dots -24^{\circ}$ in declination as center and the remaining area as periphery, we find no significant difference between them. The background density of $1.5\cdot10^{-4}\mathrm{\,arcsec}^{-2}$ is in good agreement with the value  $1.6...1.7\cdot10^{-4}\mathrm{\,arcsec}^{-2}$ that we derived from the survey of \cite{bklt} by counting the stars brighter than $\mathrm{m}_\mathrm{K}=14\mathrm{\ mag}$.

Could chance coincidences with background galaxies have led to a false classification as young stellar object or to a spurious detection of a companion? The answer is no. The galaxy counts performed with the same instrument as used by \cite{iso-oph} give values of about 25000\,sr$^{-1}$ \citep{serjeant00} for the mid-infrared and the relevant sensitivity limits of 5\,mJy at 6.7\,$\mu$m and 10\,mJy at 14\,$\mu$m. The probability is only 1\% that any one of the objects in our sample could be close enough to such a galaxy, i.e.~within 9'', to have its mid-infrared photometry affected by the presence of this galaxy. Similarly, the K-band galaxy counts \citep{gardner93, huang01} result in about 0.1 galaxies per magnitude interval per square degree at $\mathrm{m}_\mathrm{K}=10\mathrm{\ mag}$. The number increases with magnitude $\propto10^{0.67\mathrm{m}_\mathrm{K}}$ down to $\mathrm{m}_\mathrm{K}=16.5\mathrm{\ mag}$. This gives a probability of 1.4\% that any one of the objects could have the photometry affected by a close galaxy, i.e.~residing within a 4'' diameter. A probability of only 14\% is found that any of the companions within our limits of 6.4'' radius and $\mathrm{m}_\mathrm{K}\leq 14\mathrm{\ mag}$ would be a background galaxy.

\subsection{Surface Density}
\label{sd}

\begin{figure}
\centering
\includegraphics[height=8.8cm,angle=270]{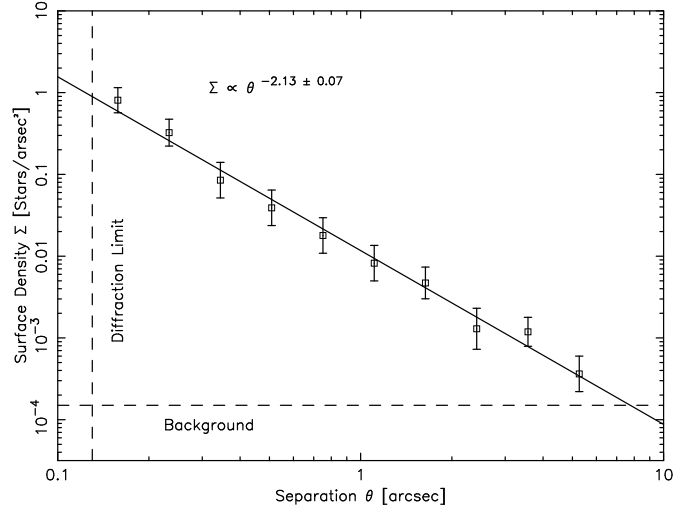}
\caption{Surface density of the companions, compared to the surface density of the background stars.}
\label{surden}
\end{figure}

An interesting property of a star forming region is the surface density $\Sigma(\theta)$ of companions (see Fig.~\ref{surden}). Over the separation range $0.13'' \leq \theta \leq 6.4''$ a linear regression of the surface density leads to
\begin{equation}
\Sigma(\theta)\propto\mathrm{\theta}^{\ -2.13\pm0.07}\mathrm{\ ,}\label{eq1} 
\end{equation}
which means that the number of companions is almost constant per logarithmic separation interval (see Fig.~\ref{results}). This is nearly the same value as derived for the Taurus star forming region \citep{koehler1}. Due to the enlarged samples, both results put the conclusion of \cite{simon} on a firmer footing, that the surface density of companions in the binary regime in different star forming regions (Taurus, Ophiuchus, Orion Trapezium) can be approximately described by $\theta^{-2}$.

The surface density of the companions is used to provide an upper limit for the separations in our survey. We choose 6.4'' (half the field of view of the SHARP cameras), because chance projections of background or foreground objects would become important at larger separations.

\subsection{Wide companions}

The field of view of the SHARP~I and the SHARP~II+ cameras is 12.8''. Usually, we centered the object in one of the lower quadrants, where the number of bad pixels was lowest. We therefore should have seen all companions out to 3.2'' in these measurements. For the wider companions we searched the 2MASS database. We found 11 infrared sources with a second source detected in the K-band within our separation range. These objects are marked with a `C' in Table~\ref{Sample}. None of them added to our list of companions (Table~\ref{Doubles}), because they had been already detected in our speckle data or were doubtful.

The `wide companions' of the sources \object{Elias~21} ($\mathrm{m}_\mathrm{K}^\mathrm{comp}=11.015~\mathrm{mag}$), \object{VSSG~18} ($12.284~\mathrm{mag}$) and \object{VSSG~17} ($13.291~\mathrm{mag}$) are indicated in the 2MASS All Sky Catalog as point sources falling within the elliptical boundary of an extended source. This suggests that the point sources are extractions of pieces of underlying nebulae. A visual inspection of the 2MASS images strengthens this suspicion. A similar case is the spurious source 5.5'' west of \object{GSS~32} with a brightness of $m_\mathrm{K}^\mathrm{comp}=13.339~\mathrm{mag}$. It is probably an artifact. Although \cite{sgl} found \object{GSS~32} single, they did not reach the necessary sensitivity to falsify the wide companion ($m_K\leq 9.2~\mathrm{mag}$). \cite{terebey} and \cite{haisch} classified \object{GSS~32} as a single star. In our fitscubes of \object{LFAM~3} \object{GSS~32} appears in the upper, i.e. eastern quadrants of the chip. No companions are visible. 

We preferred in the case of the south and the north component of \object{SR~24} the separation and position angle following from the coordinates given in 2MASS over the discrepant relative position reported in \cite{sgl} from where we adopted the values for the close pair Na-Nb.

\subsection{Number of Systems after Background Subtraction}

We find in our sample of 158 targets 49 fully resolved companions in the separation range $0.13''\leq\theta\leq 6.4''$. It thus contains 112 single stars, 43 binaries, 3 triples, and no quadruples. In addition we have to take into account that the probability $p$ to detect a background star close to a surveyed star is (section~\ref{back})
\begin{equation}
p = \pi\cdot\left(6.4\mathrm{\,arcsec}\right)^2\cdot1.5\cdot10^{-4}\mathrm{\,arcsec}^{-2}\approx0.019\label{eq2a} 
\end{equation}
or $\approx$ 3 companions in the whole sample. Therefore, three of the companions should be chance projections. This leads to a companion star frequency of $0.29\pm 0.04$. To correct the number of single, binary, and triple systems, we have to take into account that, e.g. `false' triple systems can be produced with a probability of $p$ by the `true' binaries and with a probability of $p^2$ by `true' single star. Otherwise, e.g. the number of `true' single systems is increased when compared to the number of `observed' single systems by a factor $1/(1-p+\mathcal{O}(p^2))$, because projected companions reduce their number. A brief calculation leads to 114.2 `real' single stars, 41.7 binaries, and 2.2 triple systems.

\subsection{The Restricted Sample}

For statistical purposes we also define a restricted sample, excluding all targets with uncertain association (`U' in Table~\ref{Sample}) and including only companions with brightness ratios $\geq 0.1$ where we are complete and with separations exceeding the diffraction limit. The brightness ratio of 0.1 for these young stars approximately corresponds to the limit in mass ratio of 0.1 used for the work on solar-like main-sequence stars \citep{dm}. This restricted sample contains 38 companions around 139 primaries. For the restricted sample we find 103 single stars, 34 binaries, and 2 triple systems. The background density is only $0.6\cdot10^{-4}\mathrm{\,arcsec}^{-2}$ for a detection limit of 12\,mag on average. With $p=0.008$ this sample thus contains 103.8  `real' single stars, 33.4 binaries, and 1.7 triple systems. The companion star frequency is $0.27\pm 0.04$.


\section{Discussion}

\subsection{Comparison to main-sequence stars}

\begin{figure}
\centering
\includegraphics[height=8.8cm,angle=270]{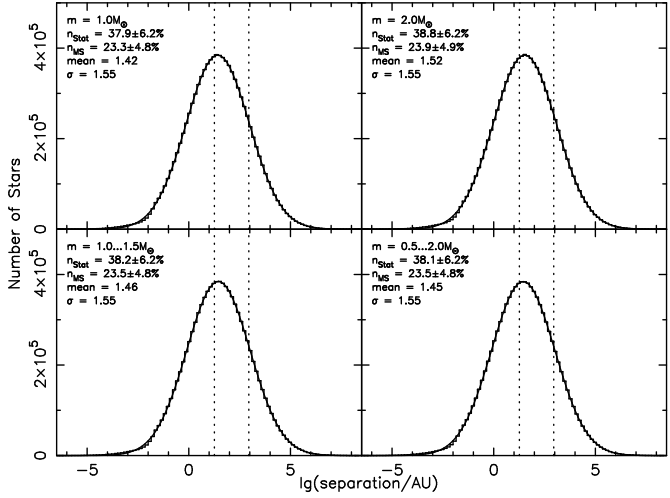}
\caption{Simulated distributions of projected separations for four samples of 10 million main-sequence binaries each with different system masses or mass ranges. The histogram shows the simulated data; the line is a lognormal distribution fitted to the histogram. The dotted vertical lines border the separation range we have observed when assuming a distance of 140~pc to the cloud complex.}
\label{ros}
\end{figure}

To compare our results with the solar-type main-sequence sample surveyed by \cite{dm} we transform their lognormal period distribution 
\begin{equation}
f\left[\lg(P)\right] = C \exp \left[-{1\over 2\sigma^2}\left(\lg(P)-\langle\lg(P)\rangle\right)^2\right]\label{eq3}
\end{equation}
with $\langle\lg(P)\rangle = 4.8$, $\sigma_P=2.3$, and $P$ in days into a lognormal distribution of separations. This is not trivial since our observations are snapshots, i.e. we cannot derive periods by fitting the orbits.  

For random distribution of orbital planes the relation between semi-major axis and actual observed separation is given by \citep{leinert93}
\begin{equation}
\langle r\rangle= {\pi\over 4}a\left(1+{e^2\over 2}\right)\mathrm{.}\label{eq4}
\end{equation}
The combined reduction of the average separation with respect to the semi-major axis would be by a factor of 0.98 if the eccentricities follow the distribution \citep{dm}
\begin{equation}
f(e) = 2e\mathrm{.}\label{eq5}  
\end{equation}
This allows to convert the orbital periods to average observed separations using Kepler's third law. With an assumed system mass of $1~M_{\odot}$ and r in astronomical units $\langle\lg(P)\rangle$ and $\sigma_P$ transform into $\langle\lg(r)\rangle = 1.48$ and $\sigma_r=1.53$. The observed separation $r$ scales with the cubic root of the total system mass. 

Alternatively, we use the well-known properties of main-sequence binaries to predict their number within the observed separation range. For Fig.~\ref{ros} we simulated a sample of $10^7$ systems with different masses or mass ranges (values in the plots). These systems have orbital elements according to \cite{dm}, i.e. the periods have the lognormal distribution (\ref{eq3}) and the distribution of eccentrities is (\ref{eq5}). The inclinations are distributed isotropically and the other parameters uniformly. After binning the results we fitted the distribution with a Gaussian (solid line). For a total system mass of $1~M_{\odot}$ we obtain $\langle\lg(r)\rangle = 1.42$ and $\sigma_r=1.55$. While $\sigma_r$ is constant for all masses and mass ranges the mean value increases as expected from lower to higher masses. For the plots in the following sections we will use $\langle\lg(r)\rangle = 1.45$ and $\sigma_r=1.55$, i.e the average of the values resulting from the system mass distributions $0.5~M_{\odot} \dots 2.0~M_{\odot}$ and $1.0~M_{\odot} \dots 1.5~M_{\odot}$. They are in good agreement with our results above.

We assume 140~pc for the distance to the Ophiuchus Dark Cloud. The separation range thus covered by our sample is marked by the vertical dotted lines in Fig.~\ref{ros}. $\mathrm{n}_\mathrm{Stat}$ is the percentage of the systems falling within these limits. After multiplying this value with the corrected \citep{dm} multiplicity of the main-sequence sample (101 companions out of 164 systems) we find $\mathrm{n}_\mathrm{MS}$, the number of companions we would have found if we had observed a sample of main-sequence stars in our survey. Due to the fact that with increasing masses the peak of the Gaussian drifts to larger separations and into our observation range, $\mathrm{n}_\mathrm{MS}$ also increases with the mass of the systems. However, for the masses considered here this effect is negligible. From the two plots with mass ranges we find
\begin{equation}
\mathrm{n}_\mathrm{MS} = (23.5\pm 4.8)\%\mathrm{.} \label{eq6}
\end{equation}
We will use this value as reference.

\begin{figure}
\centering
\includegraphics[height=8.8cm,angle=270]{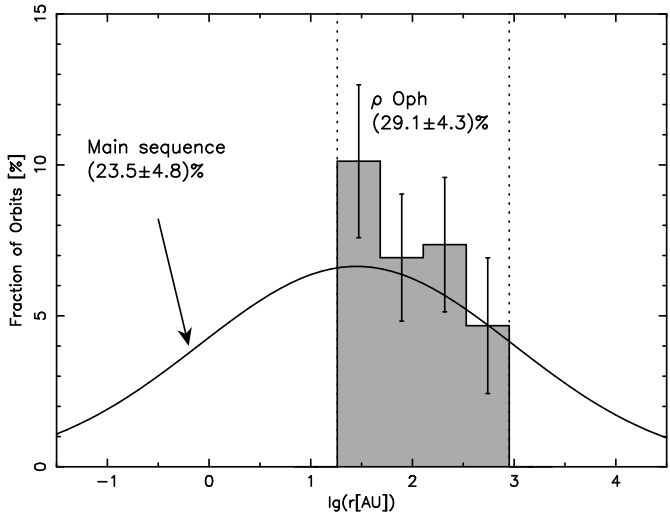}
\caption{Binary frequency as a function of separation for the total sample. The curve is the distribution of binaries among solar-type main-sequence stars \citep{dm}.}
\label{bfa}
\end{figure}

After binning all companions within the separation range $0.13''\leq\theta\leq 6.4''$ into four bins and subtracting the background we plot the result of our survey in Fig.~\ref{bfa}. Four bins are chosen since the original histogram by \cite{dm} also contains approximately four bins for the relevant range of separations or periods, respectively. The error is estimated as $\sqrt{N}$. Comparing the slope of the distribution with that of the main-sequence we find good agreement. An exception is the overabundance of close companions (see~section~\ref{miss}). With a value of 
\begin{equation}
\mathrm{n}_\mathrm{\,Oph} = (29.1\pm 4.3)\% \label{eq7}
\end{equation}
the multiplicity is only $1.24\pm 0.31$ times larger than for the main-sequence stars (\ref{eq6}). For the restricted sample (see Fig.~\ref{res}) we have
\begin{equation}
\mathrm{n}^\mathrm{\,res}_\mathrm{\,Oph} = (26.6\pm 4.4)\% \label{eq8}\\
\end{equation}
or $1.13\pm 0.30$ times the value for a main-sequence sample (\ref{eq6}).  We find the multiplicity in Ophiuchus marginally larger than for the main-sequence, but the difference is on the level of one $\sigma$ only.

\subsection{Comparison to previous surveys}

The appraisal of multiplicity among young stars `in Ophiuchus' keeps changing. From the beginning, it has been centered on a comparison to the multiplicity observed in the Taurus-Auriga star-forming region.

\cite{ghez} observed the 24 known young stars brighter than $m_\mathrm{K}=8.5\,\mathrm{mag}$ in Scorpius and Ophiuchus and found no difference with respect to Taurus in the range of separations between 16~AU and 252~AU ($0.1''-1.8''$), but a value of duplicity by a factor of 4 greater than that of the solar-type main-sequence stars (Duquennoy \& Mayor 1991). \cite{sgl} had a sample of 35 sources, selected by the randomness of lunar occultation observations (location of observatories, committee approval, weather, instrumental efficiency). Supplemented by imaging for larger separations, they found that between 3~AU and 1400~AU ($0.05''-10''$) Ophiuchus had a binary frequency of 1.1 $\pm$ 0.3 times that of nearby solar-type stars, while for Taurus this number was $1.6\pm 0.3$. These are lower limits because no corrections for incompleteness were applied. \cite{duchene} added the Gunn z band observations of \cite{rz} to these earlier surveys, corrected for incompleteness and found an enhancement of multiplicity by a factor of $1.5 \pm 0.3$ ($2.0\pm 0.3$, when \cite{sgl} is not included) over the main-sequence value, quite the same as for the Taurus-Auriga association. 

Barsony~et~al.~(2003) restricted the sample to objects searched by high-resolution near-infrared techniques. Adding new observations of this type for 19 optically selected sources from the environment of the main cloud \object{L1688}, they arrived at an overabundance of a factor of 2 $\pm$ 1 with respect to the main sequence for their sample of 80 objects, consistent with the values for the Taurus-Auriga star forming region. \cite{duchene3} did a deep ($3\mathrm{\,mag}\leq\Delta\mathrm{m}^\mathrm{lim}_\mathrm{K}\leq 7\mathrm{\,mag}$) near-infrared imaging survey of 63 embedded young stellar objects in Ophiuchus and Taurus, concluding that in the range of $110 - 1400$~AU ($0.8'' - 10''$) the multiplicity is about twice as large as for nearby solar-type main-sequence stars, and with no difference between Taurus and Ophiuchus. In this study the most embedded sources showed the highest multiplicity, still by a factor of 1.5 larger than the average. The latter result is similar to the findings in \cite{haisch} on a sample of 19 embedded objects in Ophiuchus and Serpens. Our survey, with a duplicity of young stars in Ophiuchus close to that of the main-sequence sample of Duquennoy \& Mayor, is similar in result to the study of \cite{sgl} again. 

While the studies of \cite{haisch} and \cite{duchene3}, performed on small samples, delineate interesting and important trends with age of the objects, the difference of our work to the work of \cite{barsony2} needs some explanation. As shown in the appendix, the difference will not lie in the different efficiency of the surveys, since binary young stars are consistently found in the overwhelming majority of cases by both surveys with quantitatively good agreement. Differences then should result from the selection of the sample and the angular limits over which duplicity is considered. Barsony~et~al.~(2003) in their survey and compilation of 80 objects, found $0.24 \pm 0.11$ companions per primary for the range of $0.1''-1.1''$. Choosing from their paper companions in the range of $0.13''$ to $6.4''$, as applied in our study, the resulting number of companions per primary would increase to $0.33\pm 0.07$, or $1.4\pm 0.4$ above the expectation for the main-sequence sample of solar-type stars. Otherwise, when restricting our sample to separations between $0.13''$ and $1.1''$ we find a companion star frequency of $0.16\pm0.03$. The differences are thus within the errors and naturally to be explained by differences in the samples. This just shows again the importance of large samples and to keep the sample by definition as complete as possible. The current survey with the selection criterion to take {\em all} stars brighter than $m_\mathrm{K}=10.5\,\mathrm{mag}$ that have shown convincing signs of youth compares well with previous work.

\subsection{Implications for the Formation Process}

The general frame in which we are looking at the data is the scenario that stars originally form with a high multiplicity, which then is reduced to the main-sequence value in dense environments by dynamical interactions on a short time scale. This does not mean that we want primarily to confirm this image, but that we want to check which comments or corrections with respect to this picture result from our study.

\subsubsection{Density} 

Both the Taurus-Auriga and the $\rho$ Ophiuchi molecular clouds are located at a distance of about 140~pc, contain of the order of $10^4\, \mathrm{M}_{\odot}$ of gas and dust and harbour several hundreds of young stars with an age of at most a few million years. What causes the smaller binary frequency found in our survey when compared with the result
\begin{equation}
\mathrm{n}_\mathrm{\,Tau} = (48.9\pm 5.3)\% = \left(1.93\pm 0.26\right)\mathrm{n}_{\mathrm{MS}} \label{eq9}
\end{equation}
found by \cite{koehler1} for Taurus-Auriga? $\mathrm{n}_{\mathrm{MS}}$ is the main-sequence binary fraction between their diffraction limit of 0.13'' and their upper limit of 13''. We find
\begin{equation}
\mathrm{n}^\mathrm{\,res}_\mathrm{\,Tau} = (39.7\pm 4.8)\% = \left(1.56\pm 0.31\right)\mathrm{n}_{\mathrm{MS}} \label{eq10}
\end{equation}
after all companions with a flux ratio less than 0.1 have been removed.

Taurus-Auriga is the prototypical site of low-mass star-formation. Various studies of the large-scale structure have revealed a complex, irregular, and filamentary appearance. Embedded along this filamentary structures small ($\approx$\,0.1~pc) and dense ($\ge 10^4 \mathrm{cm}^{-3}$) cores have been identified in which the young stars are forming. Their typical mass is 1~M$_{\odot}$ and their kinetic temperature about 10~K. Typical visual extinctions are between 5 and 10 mag. The whole Taurus-Auriga aggregate covers an area of 300~pc$^2$ and thus the stellar surface density is a few $\mathrm{stars\,pc}^{-2}$. Only weak clustering is apparent.

Similar conditions are found when studying the outer regions of the $\rho$ Oph complex. Loose filamentary and clumpy structures can be easily identified. A  different environment is present in the main cloud \object{L1688}. This westernmost cloud contains in an area of only $1\times 2\,\mathrm{pc}$ a centrally condensed core of $600\,\mathrm{M}_{\odot}$ with active star formation. A large fraction of all young stellar objects in the $\rho$ Oph molecular cloud are concentrated in this cluster. Stellar surface densities one or two orders of magnitudes higher than the values found in Taurus-Auriga are the result. Peak values of $5\cdot 10^3$ stars\,pc$^{-3}$  within the densest cores \citep{allen02} are almost comparable to the values that are found in the \object{Orion nebula cluster}, although there the high densities extend over larger scales. The high star-formation efficiency in \object{L1688} suggests that the cluster may remain gravitationally bound and thus survive as an open cluster \citep{lada93}. The visual extinction can reach values between 50 and 100~mag. While the radiation field caused by massive stars plays a minor or even no role for the Taurus-Auriga complex \citep{zinnecker93}, the $\rho$ Ophiuchi molecular clouds are highly influenced by the nearby Upper Scorpius-Centaurus OB-association. This may be reflected by the cometary shape of the complex and the high density reached within \object{L1688}. The influence of nearby massive stars may have also triggered the rapid rise of star-formation about 1 million years ago in the central cloud \object{L1688} \citep{palla}. All in all the $\rho$ Oph Dark Cloud with its embedded cluster seems to be an important link between loose T associations and dense clusters. 

\begin{figure}[h]
\centering
\includegraphics[height=8.8cm,angle=270]{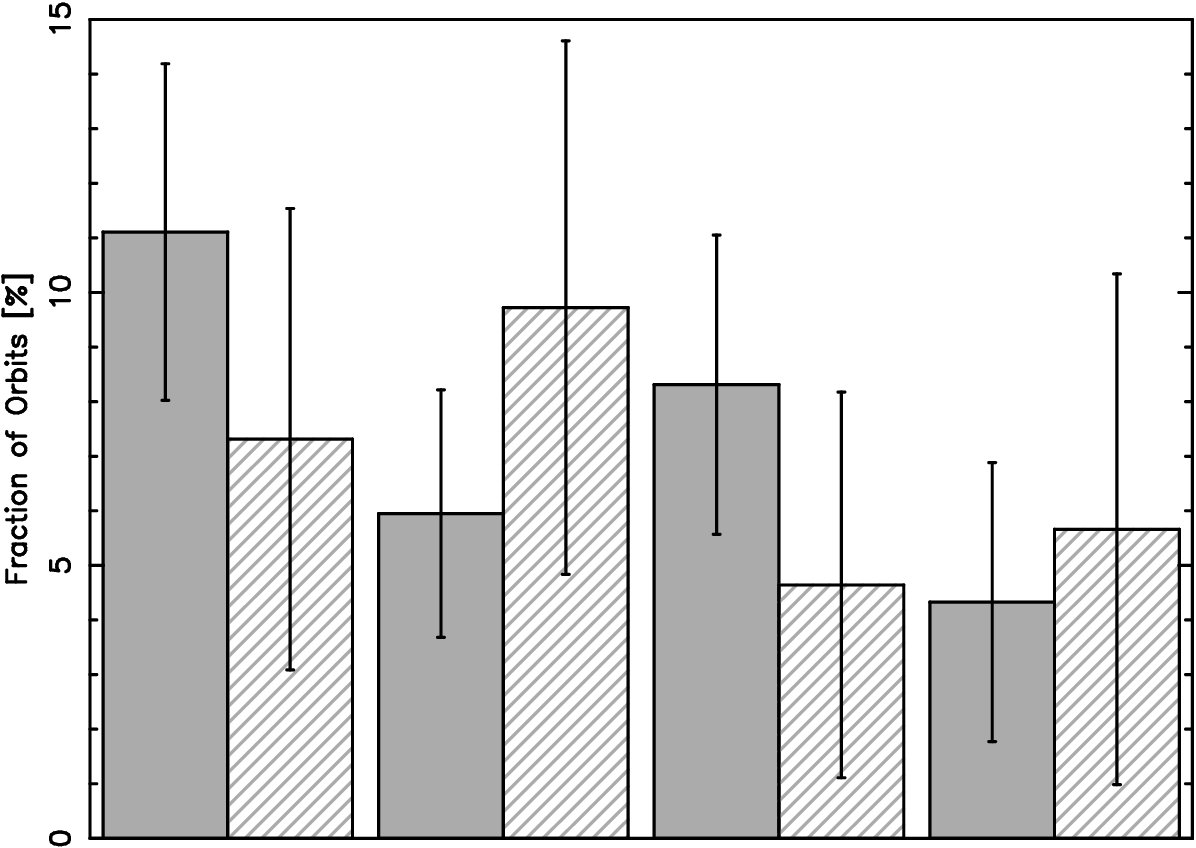}
\caption{The four bins plotted in Fig.~\ref{bfa}, but plotted for both the 117 primaries in the center (solid) and the 41 primaries in the periphery (hatched).}
\label{bfpc}
\centering
\includegraphics[height=8.8cm,angle=270]{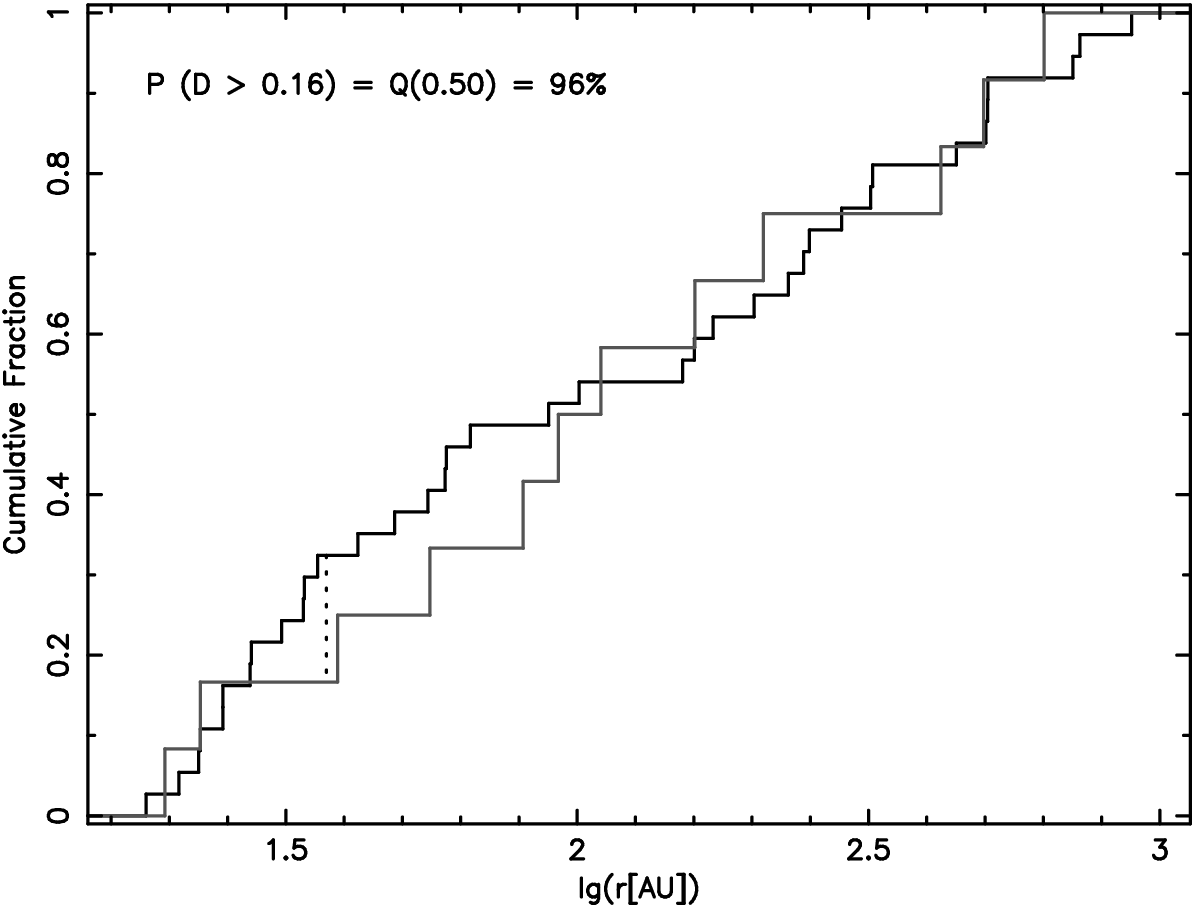}
\caption{A Kolmogorov-Smirnov test for the two datasets (center: black, periphery: grey). With a probability of 96\% the two data sets are from the same sample.}
\label{ks}
\end{figure}

\cite{duchene} in his quantitative comparison of various multiplicity surveys found that all dense clusters have binary fractions compatible with the main-sequence, while all the regions with a binary excess are loose associations. This favours a tight correlation between the density or a related parameter and the multiplicity of a star forming region. Our results seem to fit very well in this picture with a duplicity value lying between those of Taurus-Auriga and the main-sequence and hence dense clusters, both for the full and the restricted samples. In the context of dependence of duplicity on density of the star-forming region this would be a plausible result, more plausible than the large overabundance of companions found in some of the earlier surveys with smaller samples. 

One consequence of this density hypothesis would be a difference between the multiplicity of the dense central region (\object{L1688}) and that of the less dense outer regions. Recalculating the companion frequency of the total sample for both the 117 sources within \object{L1688} ($16^h25^m\dots 16^h30^m$, $-25^{\circ}\dots -24^{\circ}$) and for the 41 sources in the periphery reveals that the multiplicity of both is very similar:
\begin{eqnarray}
\mathrm{n}_\mathrm{\,Cen} & = (29.7\pm 5.0)\% & = \left(1.26\pm 0.34\right)\mathrm{n}_\mathrm{MS}\mathrm{\,,}\label{eq11}\\
\mathrm{n}_\mathrm{\,Per} & = (27.3\pm 8.2)\% & = \left(1.16\pm 0.42\right)\mathrm{n}_\mathrm{MS}\mathrm{\,.}\label{eq12}
\end{eqnarray}
Replotting the four bins displayed in Fig.~\ref{bfa} shows that the distribution of the separations of both subsamples differ only slightly from each other (Fig.~\ref{bfpc}). A Kolmogorov-Smirnov test (Fig.~\ref{ks}) also favours a common distribution. The corresponding values for the restricted sample with 104 targets in the center and 38 targets in the periphery are
\begin{eqnarray}
\mathrm{n}^\mathrm{\,res}_\mathrm{\,Cen} & = (26.2\pm 5.0)\% & = \left(1.11\pm 0.31\right)\mathrm{n}_\mathrm{MS}\mathrm{\,,}\label{eq13}\\
\mathrm{n}^\mathrm{\,res}_\mathrm{\,Per} & = (27.8\pm 8.9)\% & = \left(1.18\pm 0.45\right)\mathrm{n}_\mathrm{MS}\mathrm{\,}\label{eq14}
\end{eqnarray}
with a probability of 29\% that the two distributions are drawn from the same sample.  

This means that locally we cannot see a density effect of duplicity within the errors. The sources in the surroundings appear older on average than those within \object{L1688} and part of them could have formed in a denser environment now dissolved. However, this is nothing more than a somewhat vague possibility. Our conclusion therefore is not as clear as one might want it to be.

{\it Although density, or a related parameter seems to play a crucial role in the formation of binaries on a global scale, there is no statistical significance within the $\rho$ Oph molecular cloud complex that areas with different densities show different multiplicities.}

\subsubsection{Temporal Evolution} 

Star forming regions with a main-sequence binary fraction are found at all ages, e.g. IC~348 \citep{duchene2}, Orion \citep{petr} with an age of a few million years, the Pleiades \citep{bouvier1} with 120~Myr, and the Praesepe \citep{bouvier2} with 700~Myr. This suggests that dynamic interactions, if responsible for reducing an originally high duplicity to much lower values, act very quickly in dense clusters, while little future effect has to be expected for low-density regions like Taurus-Auriga. Thus temporal evolution of the binary frequency is not in general responsible for the difference between the overabundance of companions found in Taurus-Auriga when compared to the main-sequence. The fact that in the young but not too dense Ophiuchus star-forming region there remains an overabundance of companions, with the most embedded sources showing the highest degree of multiplicity \citep{duchene3}, would be compatible with the dynamical evolution of binarity in cluster environments \citep{kroupa}.

\begin{figure}
\centering
\includegraphics[height=8.8cm,angle=270]{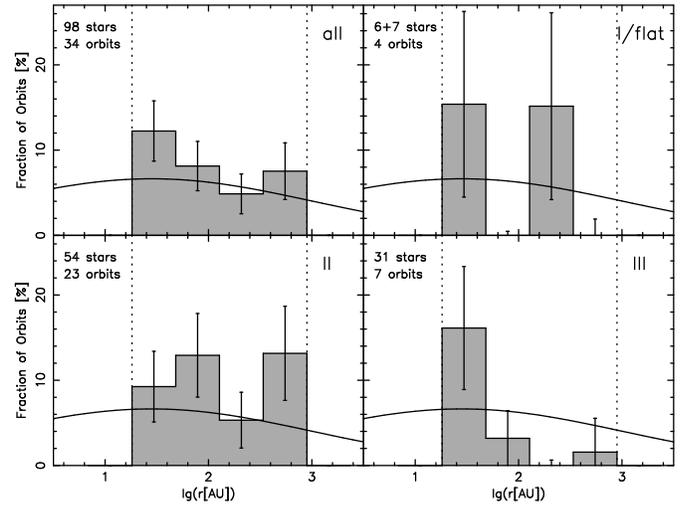}
\caption{Binary frequency as a function of separation and class. The upper left panel shows the distribution for all classified sources. The remaining panels display the combined sample of flat spectrum and class~I sources, the class~II, and the class~III samples. The curve is the distribution of binaries among solar-type main-sequence stars \citep{dm}.}
\label{bfclass}
\end{figure}

We tried to see this effect in our sample with respect to the age of the different objects. We thus searched for their infrared classes in the literature. Although this classification scheme is more a morphological description than a direct indicator of the age, it provides the best approach when no spectroscopic data are available. To avoid systematic errors from different surveys we only used the classification provided by the mid-infrared survey of \cite{iso-oph} and the near-infrared study of \cite{gwayl}. To be consistent with the classification in \cite{iso-oph} we decided to classify in \cite{gwayl} objects with a spectral slope $a > 0.55$ as class~I and those with $a > -0.05$ as flat spectrum sources. To distinguish between more evolved class~II and class~III objects in \cite{gwayl} we used $a = -1.6$ as limit. \object{WL~5} is an exception, since it is classified as an heavily reddened class~III source. This conclusion is in agreement with the result in \cite{iso-oph}. All sources included in both samples are classified consistently with exception of \object{L1689-IRS~5} and \object{LFAM~3} that are according to \cite{gwayl} flat spectrum sources, but are classified as class~II sources in \cite{iso-oph}. Since \object{LFAM~3} lies only marginally below the limit in \cite{iso-oph} we decided to classify it as flat spectrum source. Otherwise, \object{L1689-IRS5} is only slightly above the limit in \cite{gwayl} and well below in \cite{iso-oph}. We thus classified it as class~II object. Class~I and class~II sources in \cite{gwayl} with an upper limit for $a$ are ignored.  This leads to a sample of 6 class~I, 7 flat spectrum, 54 class~II, and 31 class~III sources. The multiplicity of this subsample is
\begin{equation}
\mathrm{n}_\mathrm{I-III} = (32.8\pm 5.8)\% = \left(1.39\pm 0.38\right)\mathrm{n}_\mathrm{MS} \label{eq15}
\end{equation}
and thus compatible with the result found in (\ref{eq7}) and (\ref{eq8}) within the error bars. After separating the different evolutionary states we are left with subsamples that are no longer free from small number statistics (see Fig.~\ref{bfclass}): 
\begin{eqnarray}
\mathrm{n}_\mathrm{I/flat}   & = (29\pm 15)\%  & = \left(1.2\pm 0.7\right)\mathrm{n}_\mathrm{MS} \mathrm{\,,}\label{eq16}\\
\mathrm{n}_\mathrm{II}       & = (41\pm 9)\%  & = \left(1.7\pm 0.5\right)\mathrm{n}_\mathrm{MS}  \mathrm{\,,}\label{eq17}\\
\mathrm{n}_\mathrm{III}      & = (21\pm 8)\%   & = \left(0.9\pm 0.4\right)\mathrm{n}_\mathrm{MS} \mathrm{\,.} \label{eq18}
\end{eqnarray}
There appears to be a trend that class~III systems (WTTS) have fewer companions and at smaller separations than their class~II (CTTS) counterparts. This was not found in Taurus. \cite{ghez} suggested from a similar result on a smaller sample that close companions may help to clear circumstellar disks earlier and therefore appear more frequently in WTTS.

Temporal evolution may be important in dense environments at early stages. In our sample of stars located in a cluster of medium density we are less sensitive to such an effect. However, the difference in the multiplicity and separation distribution between class~II and class~III sources and with respect to Taurus could nevertheless show real changes, maybe temporal evolution.

Another possibility is a biasing of the sample by a yet not distinguished older population of lower multiplicity. In the last section we excluded a strong influence of such a population in the periphery. Nevertheless, if the stars reside instead in the foreground, they could mimic the here discussed difference between the classes. Precise measurements, e.g.~with GAIA of the parallaxes will test this idea. 

{\it Although temporal evolution seems to be not responsible for the reduction of the binary frequency in general except for the earliest stages, our survey indicates statistical differences between the infrared classes with respect to their companion frequencies and separation distributions. }

\subsubsection{Missing Companions}
\label{miss}

Two possible explanations have been discussed by \cite{duchene} for a low multiplicity of the Ophiuchus star forming region when compared to Taurus-Auriga. a) The distribution of the projected separations can be shifted to lower values, i.e. the `missing' companions are {\it too close} to be resolved and are hidden from our survey below the diffraction limit. b) The flux ratio of the companions is smaller for Ophiuchus, i.e. the `missing' companions are {\it too faint} to be detected. 

\begin{figure}[h]
\centering
\includegraphics[height=8.8cm,angle=270]{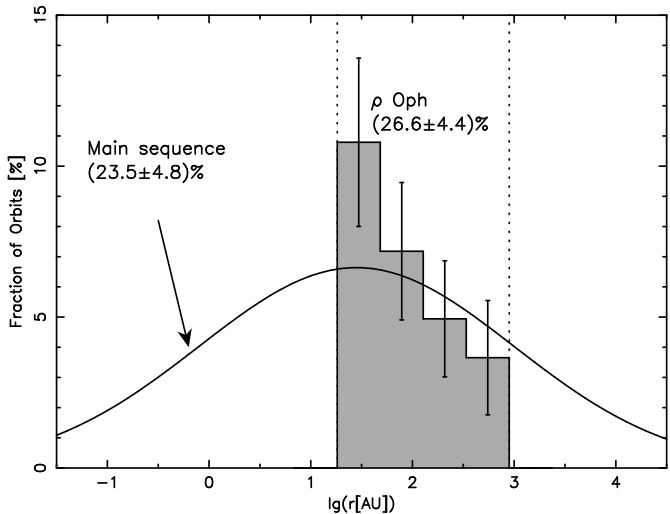}
\caption{Binary frequency as a function of separation for the restricted sample. The curve is the distribution of binaries among solar-type main-sequence stars \citep{dm}.}
\label{res}
\end{figure}

To conclude on the first possibility, very high resolution observations (lunar occultations or interferometry) would have to be available for most of the sources of our survey, which is not yet the case. From \cite{sgl} and \cite{barsony2} there is at least evidence that no overabundance of stars with very close companions is present. However, both studies suffer from poor statistics.  Fig.~\ref{res} that displays the multiplicity as function of the separation for our restricted sample shows a trend that the sample is dominated by close companions. This overabundance is more apparent in the restricted sample than in the total sample (Fig.~\ref{bfa} and Fig.~\ref{res}).

\begin{figure}[h]
\centering
\includegraphics[height=8.8cm,angle=270]{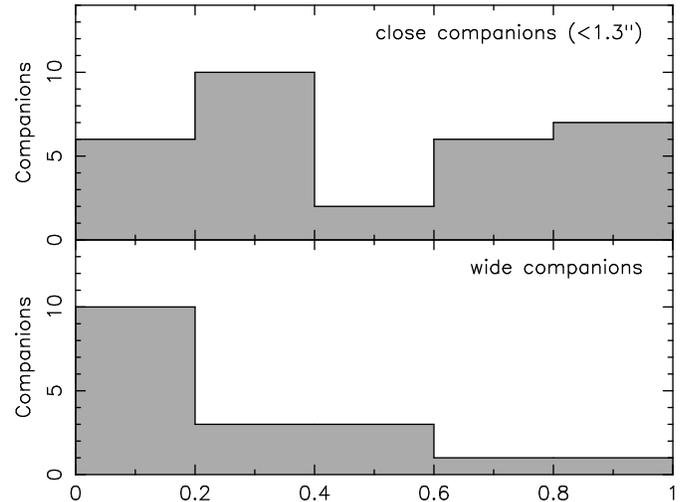}
\caption{Flux ratio for close ($<$1.3'') and wide companions.}
\label{fr}
\end{figure}

Concerning the second suggestion, \cite{duchene} found from \cite{ghez} that 73\% of the binaries in Taurus, but only 23\% of the binaries in Ophiuchus exhibit a magnitude difference between companion and primary of $\Delta\mathrm{m}_\mathrm{K} < 1.5\,\mathrm{mag}$. We want to check whether this also holds for our larger sample. In Fig.~\ref{fr} the flux ratios of our companions in the total sample are displayed. Indeed, the whole sample is dominated by small flux ratios. About 59\% of the systems show flux ratios below 0.4 and 33\% below 0.2. This tendency to favour small flux ratios is introduced by the wide pairs ($>$1.3''). The close companions are almost equally distributed. Even when the boundary between close and wide companions is varied, Kolmogorov-Smirnov tests show that the probability that the two distributions have a common origin is below 10\%.

In \cite{koehler1} wide pairs ($>$1.3'') are also dominated by small flux ratios, similar to the result displayed in Fig.~\ref{fr}. On the other hand there is a clear tendency in Taurus-Auriga for close binaries to exhibit a large fraction of equally bright systems possibly caused by a lack of close binaries with small flux ratios that are present in Ophiuchus. Such a population may be the reason for the finding in \cite{duchene}.

{\it A combination of the two trends, i.e. the high fraction of close binaries and the presence of close companions with low flux ratios, leads to the conclusion that `missing' companions may play a role with the implication that the full binary fraction over all separations would be more clearly enhanced than the binarity in our restricted sample.}


\section{Summary}

\begin{itemize}

\item We presented a volume-limited multiplicity survey with magnitude cutoff ($\mathrm{m}_\mathrm{K}\leq 10.5\mathrm{\ mag}$) of 158 young stellar objects located within or in the vicinity of the $\rho$ Ophiuchi Dark Cloud (\object{L1688}). The survey covers separations between 0.13'' (diffraction limit) and 6.4'' (background contamination) and is complete for flux ratios $\geq 0.1$ ($\Delta\mathrm{m}_\mathrm{K}\leq 2.5$) at the diffraction limit. A restricted sample has been defined that is complete and excludes all uncertain cloud members.
\item The detection limit is $\mathrm{m}_\mathrm{K}\approx 14\mathrm{\ mag}$, and the stellar background density at this brightness is $\approx 1.5\cdot10^{-4}\mathrm{\,arcsec}^{-2}$
\item Among the 147 targets newly observed with speckle techniques in the K-band we found 48 companions (40 binary and 4 triple systems). Five of these companions are below the diffraction limit of the telescopes and thus only marginally resolved. From the remaining 43 companions (39 binary and 2 triple systems) 14 are new detections including a third component in the previously known binary system \object{ROXs~42B} and the resolution of the previously known companion of \object{L1689-IRS~5} into two sources.
\item The surface density of the companions $\Sigma$ as a function of the separation $\theta$ can be well fitted by the power law $\Sigma(\theta)\propto\mathrm{\theta}^{\ -2.13\pm0.07}$.
\item Within the range $0.13''\leq\theta\leq 6.4''$ our multiplicity is $(29.1\pm 4.3)\%$ for the total and $(26.6\pm 4.4)\%$ for the restricted sample.
\item This value is  $1.24\pm 0.31$, respectively $1.13\pm 0.30$ times the main-sequence value. The close similarity between Taurus and Ophiuchus found in most previous surveys is questioned by our result, which is based on a larger and more complete sample. 
\item The idea that the observed duplicity in star-forming regions is governed by some process related to the density of the stellar environment gets global support from our observations. This process has been suggested earlier to be related either to the formation process or to dynamical interaction afterwards. Observations like those of \cite{duchene} and \cite{haisch} tend to favour the second scenario. Our data are not sensitive to this alternative.
\item There seems to be a relation between spectral classes and binary fraction. Class~II objects have a multiplicity twice that of class~III objects. This relation has not been found in the Taurus-Auriga survey \citep{koehler1}.
\item Our results find their place in the paradigm of originally very high multiplicity of young stellar objects that then is reduced by dynamical interactions to different degrees in environments of different densities. This may be the global picture, however, locally within our sample we see no significant difference between the $\rho$ Ophiuchi Dark Cloud (\object{L1688}) and its less dense environment. Only the differences between class~II and class~III sources may point to evolution. 
\item A population of close binaries with low flux ratios not present in Taurus, for which first indications exist, may be a partial answer to the question why the multiplicity in Taurus-Auriga is so clearly higher than in Ophiuchus.

\end{itemize}


\appendix

\section{Comparison with previously known binary and multiple systems}

\begin{table*}
\caption[]{Comparison to previously known binaries}
\label{Barsony}$$
\begin{array}{p{0.11\linewidth}p{0.11\linewidth}p{0.11\linewidth}p{0.06\linewidth}p{0.08\linewidth}p{0.08\linewidth}p{0.06\linewidth}p{0.06\linewidth}p{0.06\linewidth}p{0.15\linewidth}}
\hline
\noalign{\smallskip}
Object & Ref & Date & PA$_{\mathrm{Lit}}$ & $\theta_{\mathrm{Lit}}$ & K$_2$/K$_{1\mathrm{, Lit}}$ & PA    & $\theta$ & K$_2$/K$_1$ & Remark\\
       &     &      & [deg]               & ['']                  &                       & [deg] & ['']     &      & \\

\noalign{\smallskip}
\hline
\noalign{\smallskip}
H$\alpha$ 18  & {\bf K}, R1	     & 1999, June 1/2  & 339.55$^\mathrm{m}$ & $\approx0.1004^\mathrm{d}$ & $\approx 0.505$   & -	 & -		      & -     & not detected\\
	      &	                     &	               & \ \ 80.4	     & 1.08			  & 0.7 	      & \ \ 82.3 & 1.083	      & 0.737 & \\    
H$\alpha$ 19  & {\bf K}, R1	     & 1999, June 1/2  & 260.7  	     & 1.53			  & 0.5 	      & 262.9	 & 1.491	      & 0.462 & \\
SR 2	      & A2, {\bf G2}         & 1990, July 8    & 156		     & 0.236			  & 0.80	      & 122.4	 & 0.222	      & 0.874 & ROX 1\\
ROXs 2	      & {\bf B2}, C	     & 2002, May 24    & 347.1  	     & 0.42			  & 0.57	      & 345.5	 & 0.424	      & 0.598 & \\
IRS 2	      & {\bf B2}, C	     & 2002, May 24    & \ \ 77.6	     & 0.42			  & 0.13	      & \ \ 78.6 & 0.426	      & 0.132 & \\
ROXs 5	      & {\bf A1}	     & 1993 / 1994     & $\approx 130$       & $\approx 0.13$		  & $\approx 0.5$     & 327.3	 & 0.176	      & 0.408 & obs. in H-band\\
H$\alpha$ 26  & {\bf K}, R1	     & 1999, June 1/2  & \ \ 23.8	     & 1.15			  & 0.9 	      & \ \ 25.8 & 1.135	      & 0.846 & \\
H$\alpha$ 28  & {\bf R1}	     & 1991 / 1992     & 358		     & 5.1			  & 0.06	      & 357.8	 & 5.209	      & 0.047 & obs. with Gunn z\\
VSSG 27	      & {\bf C}	             & 1995 / 1996     & \ \ 68 	     & 1.22			  & 0.24	      & \ \ 66.8 & 1.222	      & 0.244 & \\
H$\alpha$ 35  & {\bf K}, R1	     & 1999, June 1/2  & 130.3  	     & 2.29			  & 0.3 	      &  132.2   & 2.277	      & 0.272 & \\
GSS37	      & C, {\bf K}, R1       & 1999, June 1/2  & \ \ 67.0	     & 1.44			  & 0.3 	      & \ \ 69.5 & 1.438	      & 0.299 & \\
ROXs 16	      & A1, {\bf C}	     & 1995 / 1996     & \ \ \ \ -	     & -			  & -		      & \ \ 24.2 & 0.098$^\mathrm{d}$ & 0.357 & VSS 27, triple?\\
	      &	                     &	               & 106		     & 0.57			  & 0.06	      & 105.4	 & 0.577	      & 0.186 & \\
WL 18	      & {\bf B1}	     & 1988, June 2/3  & 293		     & 3.55			  & 0.2 	      & 292.4	 & 3.617	      & 0.162 & rel. pos. from B2\\
VSSG 3	      & {\bf C}	             & 1995 / 1996     & \ \ 47 	     & 0.25			  & 0.38	      & \ \ 53.8 & 0.243	      & 0.801 & \\
Elias 30      & M1, {\bf S2}         & 1992, June 13   & 175		     & 6.700			  & 0.030	      & 175.6	 & 6.388	      & 0.063 & SR 21\\
WL 20	      & {\bf R2}	     & 1990 / 1998     & 270.1  	     & 3.17			  & 0.70$^\mathrm{n}$ & 269.9	 & 3.198	      & 0.877 & phot. 1990, pos. 1998\\
	      &	                     &	               & 232.2  	     & 3.66			  & 0.07$^\mathrm{n}$ & 232.3	 & 3.619	      & 0.071 & in the mid-IR\\
VSSG 25	      & {\bf C}	             & 1995 / 1996     & 356		     & 0.46			  & 0.5 	      & 173.3	 & 0.468	      & 0.887 & WL 13\\
IRS 44      & C, H, S2, {\bf T}    & 1997	       & \ \ 81 	     & 0.27			  & -		      & 246.6	 & 0.256	      & 0.204 & \\
VSSG 17	      & {\bf C}	             & 1995 / 1996     & 269		     & 0.25			  & 0.2 	      & 260.2	 & 0.242	      & 0.644 & \\
SR 9	      & B2, G1, {\bf G2}     & 1990, July 9    & 350		     & 0.59			  & 0.09	      & 353.3	 & 0.638	      & 0.057 & \\
VSSG 14	      & {\bf S2} 	     & 1992, June 13   & \ \ 89$^\mathrm{p}$ & 0.101$^\mathrm{d,p}$	  & 0.5 	      & \ \ 83.6 & 0.130	      & 0.296 & \\
ROXs 31	      & A1, C, S1, {\bf S2}  & 1986 / 1991     & 262		     & 0.480			  & 0.8 	      & 251.3	 & 0.396	      & 0.655 & \\
ROXs 42B      & {\bf S2}	     & 1992, June/July & \ \ 89$^\mathrm{p}$ & 0.056$^\mathrm{d,p}$	  & 0.8 	      & 157.9	 & 0.083$^\mathrm{d}$ & 0.350 & new triple system\\
	      & 	             &		       & \ \ \ \ -	     & -			  & -		      & 268.0	 & 1.137	      & 0.002 & \\
ROXs 42C      & B2, {\bf G2}, M2     & 1990, July 8    & 135		     & 0.157			  & 0.25	      & 151.0	 & 0.277	      & 0.220 & \\
ROXs 43 A/B   & A1, M1, R1, {\bf S2} & 1992, July 11   & \ \ \ \ 7	     & 4.800			  & 0.44	      & \ \ 11.9 & 4.523	      & 0.445 & \\
H$\alpha$ 71  & {\bf K}, S2	     & 1999, June 1/2  & \ \ 35.0	     & 3.56			  & 0.17	      & \ \ 35.0 & 3.560	      & 0.151 & calibrator\\
L1689 - IRS 5 & {\bf H}	             & 2001, July 11   & 240.3  	     & 2.92			  & 0.61	      & 241.2	 & 3.006	      & 0.277 & new triple system\\
              & 	             &      	       & \ \ \ \ -	     & -			  & -		      & \ \ 84.4 & 0.140	      & 0.946 & \\
DoAr 51	      & {\bf B2}	     & 2002, May 24    & \ \ 80.8	     & 0.79			  & 0.2 	      & \ \ 79.3 & 0.784	      & 0.228 & ROXs 47A\\
\noalign{\smallskip}
\hline
\noalign{\smallskip}
\multicolumn{2}{l}{$A1: \cite{ageorges}$} & \multicolumn{3}{l}{$A2: \cite{aitken}$} 	& \multicolumn{3}{l}{$B1: \cite{barsony1}$} 	& \multicolumn{2}{l}{$B2: \cite{barsony2}$} \\ 
\multicolumn{2}{l}{$C: \cite{costa}$} 	  & \multicolumn{3}{l}{$G1: \cite{geoffray}$} 	& \multicolumn{3}{l}{$G2: \cite{ghez}$} 	& \multicolumn{2}{l}{$H: \cite{haisch}$} \\ 
\multicolumn{2}{l}{$K: \cite{koresko}$}   & \multicolumn{3}{l}{$M1: 2MASS$} 		& \multicolumn{3}{l}{$M2: \cite{mathieu}$} 	& \multicolumn{2}{l}{$R1: \cite{rz}$} \\ 
\multicolumn{2}{l}{$R2: \cite{ressler}$}  & \multicolumn{3}{l}{$R3: \cite{richichi94}$}	& \multicolumn{3}{l}{$S1: \cite{simon87}$} 	& \multicolumn{2}{l}{$S2: \cite{sgl}$} \\
\multicolumn{2}{l}{$T: \cite{terebey}$} \\ 
\noalign{\smallskip}
\multicolumn{2}{l}{^\mathrm{d}$: below our diffraction limit$} & \multicolumn{3}{l}{^\mathrm{m}$: mod 180 deg$} & \multicolumn{3}{l}{^\mathrm{n}$: non-photometric conditions$} & \multicolumn{2}{l}{^\mathrm{p}$: projected values$} \\ 
\noalign{\smallskip}
\hline
\end{array}
$$
\end{table*}

\cite{barsony2} compiled a list of all known binary and multiple systems associated with the $\rho$ Ophiuchi Dark cloud. In Table~\ref{Barsony} we list all sources of this compilation with a separation in the range $0.05''\leq\theta\leq 6.7''$ that have been newly observed within the scope of our survey. The first column gives the name used in our survey. If \cite{barsony2} used another designation it appears in the last column. The second column gives their references. The references printed in bold font are those that we used to derive the values for the position angles, the separations and the flux ratios in the columns 4, 5, and 6. The date of observation in this reference is given in column 3. For comparison our results are listed in the subsequent columns.

\begin{figure}[h]
\centering
\includegraphics[width=8.8cm]{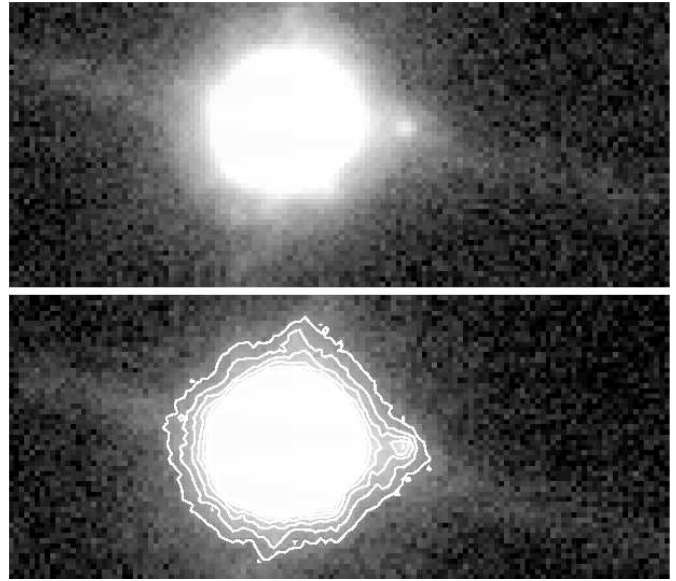}
\caption{The faint companion of ROXs~42B as seen in the shift-and-add images of the two fitscubes. The images are printed in logarithmic scale. Linear contours are overlayed on the second image. The flux ratio is 0.002.}
\label{ROXs42B}
\end{figure}

\begin{figure}[h]
\includegraphics[width=8.8cm]{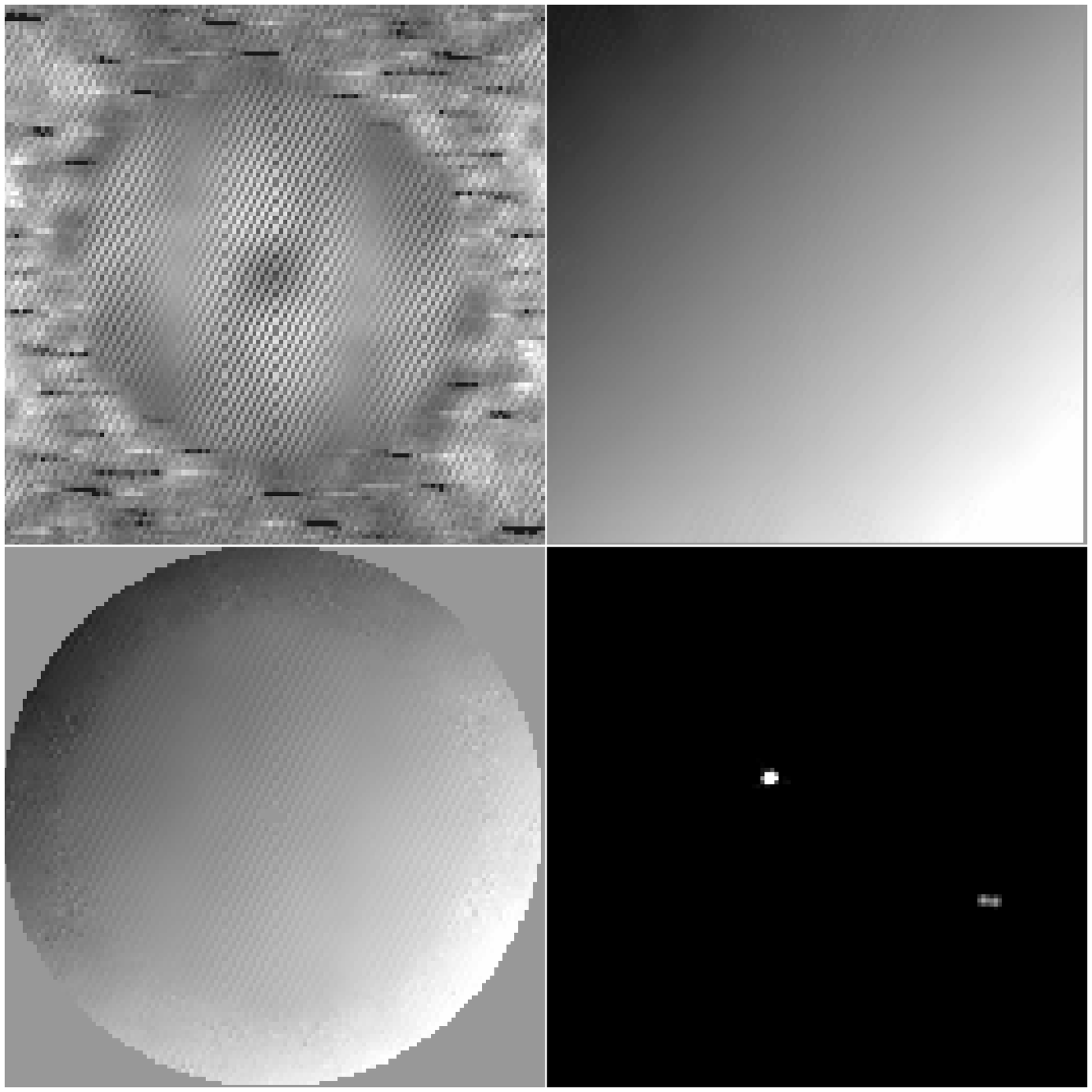}
\caption{The newly detected triple system L1689-IRS~5. Diplayed are the visibility, the Knox-Thompson phase, the bispectrum phase, and the reconstructed image. The visibility clearly shows that the wide pair (narrow stripes) is superimposed by a close pair almost at the diffraction limit (the two wide stripes).}
\label{L1689-IRS5}
\end{figure}

Only one companion was not detected in our survey. This is probably caused by the small separation below the diffraction limit of our telescope. Otherwise, we detected additional companions of \object{ROXs~42B} and \object{L1689-IRS~5} transforming these objects into triple systems. The new companion of \object{ROXs~42B} (Fig.~\ref{ROXs42B}) is a faint knot west of the primary. In the case of \object{L1689-IRS~5} (Fig.~\ref{L1689-IRS5}) the previously known companion splits up into two point sources with equal fluxes. They are visible as a elongated structure in the shift-and-add images and are very prominent in the visibility. Unfortunately, a probable third component of \object{ROXs~16} is too close to distinguish between an elongated structure and a point source.


Although the binary parameters derived in our analysis are in general very similar to those provided by the papers used as references in Table~\ref{Barsony}, some important differences exist. The position angle of the binaries \object{ROXs~5}, \object{VSSG~25} and \object{IRS~44} have changed approximately $180^{\circ}$ since the last observations. Due to the fact that the separations did not change either the flux ratio has changed significantly or the measurements suffer from the $180^{\circ}$ ambiguity. Furthermore, some flux ratios have changed at least by a factor of two: \object{ROXs~16} (maybe mainly an effect of the probable new companion), \object{VSSG~3}, \object{Elias 30}, \object{VSSG~17}, and of the close companion of \object{ROXs~42B}. Finally, the relative positions of the companions orbiting \object{SR~2} and \object{ROXs~42C} have changed significantly.

\section{Additional Sources}

\begin{table*}
\caption[]{Additional sources}
\label{A1}$$
\begin{array}{p{0.17\linewidth}p{0.09\linewidth}p{0.08\linewidth}p{0.05\linewidth}p{0.10\linewidth}p{0.04\linewidth}p{0.04\linewidth}p{0.35\linewidth}}
\hline
\noalign{\smallskip}
Object 	 & \multicolumn{2}{c}{J2000.0\ \ \ }	              & $\mathrm{K}_{\mathrm{2MASS}}$ & Date 	      & 0.15'' & 0.50'' & Remarks\\
	 & \hspace{0.5cm}$\alpha$ 	& \hspace{0.6cm}$\delta$ 	& $\mathrm{[mag]}$              &               &        &        &\\
\noalign{\smallskip}
\hline
\noalign{\smallskip}
\object{GY 45}               & 16 26 29.98 & -24 38 42.8 & \ 8.367 & 2000, June 17 & 0.04 & 0.03 & binary, background giant \citep{luhman}\\
\object{GY 65}               & 16 26 32.91 & -24 36 26.4 & \ 8.996 & 2000, June 17 & 0.08 & 0.05 & background giant \citep{luhman}\\
\object{BKLT J162637-241602} & 16 26 37.13 & -24 15 59.9 &  10.757 & 2001, July 3  & 0.15 & 0.08 & too faint\\
\object{VSS 28}	     & 16 26 52.80 & -23 43 12.7 & \ 6.702 & 2001, July 4  & 0.07 & 0.04 & binary, background determination, not a member\\
\object{VSSG 6}              & 16 26 53.86 & -24 22 28.0 & \ 9.827 & 2000, June 21 & 0.09 & 0.04 & background giant \citep{luhman}\\
\object{GY 232}              & 16 27 13.26 & -24 41 33.7 & \ 9.592 & 2000, June 21 & 0.10 & 0.07 & background giant \citep{luhman}\\
\object{VSSG 13}             & 16 27 46.69 & -24 23 22.1 & \ 7.270 & 2000, June 21 & 0.03 & 0.02 & field star \citep{elias}\\
\object{GY 411}              & 16 27 57.89 & -24 37 49.0 & \ 9.560 & 2000, June 21 & 0.07 & 0.03 & background giant \citep{luhman}\\
\object{VSSG 16}             & 16 28 03.73 & -24 26 32.0 & \ 6.504 & 2000, June 17 & 0.04 & 0.03 & field star \citep{elias}\\
\object{VSSG 15}             & 16 28 09.23 & -24 23 20.7 & \ 7.140 & 2000, June 21 & 0.07 & 0.03 & field star \citep{elias}\\
\object{ROXs 47B} 	     & 16 32 23.28 & -24 40 18.5 & \ 8.581 & 2000, June 22 & 0.04 & 0.04 & foreground \citep{ba}\\
\object{DoAr 58}             & 16 34 26.70 & -24 13 43.7 & \ 7.707 & 2001, July 4  & 0.15 & 0.04 & \object{HBC~269}, deleted in \cite{hbc}\\
\noalign{\smallskip}
\hline
\end{array}
$$
\end{table*}

During our speckle observations we recorded 12 sources not part of the final sample (see Table~\ref{A1}). \object{BKLT J162637-241602} has been observed by chance due to its location close to \object{BKLT~ J162636-241554}. Although \object{VSS~28} is probably not a member of the Ophiuchus molecular clouds, we observed it with speckle techniques, because the area around \object{VSS~28} has been used to determine the background density. While \object{DoAr~58} was listed as \object{HBC~269} in \cite{hrc}, it has been removed in the third edition of this catalogue \citep{hbc}. The remaining objects are either background giants or foreground dwarfs. 

\object{VSS~28} is a binary that has been observed with a separation of $0.344''\pm 0.005''$ at a position angle of $308.3^{\circ} \pm 0.7^{\circ}$ in the night following the 4th July 2001. The flux ratio of the two components is $0.203 \pm 0.026$ and we can exclude at a distance of 0.15'' (0.50'') from the main component companions with flux ratios larger than 0.07 (0.04). Another binary is \object{GY~45}. The 2MASS images of this source show a very symmetric extension to the south that has been detected in all three wavelength bands. We inspected our shift-and-add images and found a shallow glow at the calculated position only 2$\sigma$ above the background. The parameters of this binary can be derived from the 2MASS survey. The companion resides at a position angle of $168.4^{\circ} \pm 1.8^{\circ}$ with a separation of  $4.977''\pm 0.144''$ and a flux ratio of $0.017 \pm 0.003$. In our speckle data obtained at the 17th June 2000 we can exclude at a distance of 0.15'' (0.50'')  from the main component companions brighter than 0.04 (0.03) times the flux of the primary. All the remaining ten sources are single stars.


\begin{acknowledgements}
      We thank the staff at La Silla and at Calar Alto for their 
      support during several observing runs and Andreas Eckart and
      his team for friendly cooperation during the runs with SHARP~I
      at the NTT. We also like to thank the referee for the helpful
      comments.
      
      This publication makes use of data products from the Two 
      Micron All Sky Survey (2MASS), which is a joint project of 
      the University of Massachusetts and the Infrared Processing 
      and Analysis Center/California Institute of Technology, 
      funded by the National Aeronautics and Space Administration 
      and the National Science Foundation.      
\end{acknowledgements}


\bibliographystyle{bibtex/aa}
\bibliography{astro}


\end{document}